\documentclass[a4paper,11pt]{article}

  \usepackage{amsfonts,amsthm,amssymb}
  \usepackage{latexsym, color,epsf}

\newcommand{\eq}{\begin{equation}}
\newcommand{\en}{\end{equation}}
\newcommand{\eqa}{\begin{eqnarray}}
\newcommand{\ena}{\end{eqnarray}}

\begin{document}

  \setlength{\unitlength}{1mm}

  \thispagestyle{empty}

 \begin{center}
  {\LARGE \bf  Permutation and Its Partial Transpose}

  \vspace{.3cm}

  Yong Zhang${}^{ad}$\footnote{yzhang@nankai.edu.cn},
  Louis H. Kauffman${}^{b}$\footnote{kauffman@uic.edu}
  and Reinhard F. Werner${}^{c}$\footnote{r.werner@tu-bs.de}\\[.2cm]

 ${}^a$  Theoretical Physics Division, Chern Institute of Mathematics \\
  Nankai University, Tianjin 300071, P. R. China\\[0.1cm]

 ${}^b$  Department of Mathematics, Statistics and Computer Science\\
 University of Illinois at Chicago, 851 South Morgan Street\\
 Chicago, IL, 60607-7045, USA \\[0.1cm]

 ${}^c$ Institut f\"{u}r Mathematische Physik, TU Braunschweig,\\
 Mendelssohnstr. 3, 38304 Braunschweig, Germany \\[0.1cm]

  ${}^d$ Institute of Theoretical Physics, Chinese Academy of Sciences\\
 P. O. Box 2735, Beijing 100080, P. R. China \\[0.2cm]

\end{center}

\vspace{0.1cm}

\begin{center}
\parbox{13.5cm}{
\centerline{\small  \bf Abstract}  \noindent

Permutation and its partial transpose play important roles in
quantum information theory. The Werner state is recognized as a
rational solution of the Yang--Baxter equation, and the isotropic
state with an adjustable parameter is found to form a braid
representation. The set of permutation's partial transposes is an
algebra called the ``PPT" algebra which guides the construction of
multipartite symmetric states. The virtual knot theory having
permutation as a virtual crossing provides a topological language
describing quantum computation having permutation as a swap gate. In
this paper, permutation's partial transpose is identified with an
idempotent of the Temperley--Lieb algebra. The algebra generated by
permutation and its partial transpose is found to be the Brauer
algebra. The linear combinations of identity, permutation and its
partial transpose can form various projectors describing tangles;
braid representations; virtual braid representations underlying
common solutions of the braid relation and Yang--Baxter equations;
and virtual Temperley--Lieb algebra which is articulated from the
graphical viewpoint. They lead to our drawing a picture called the
``ABPK" diagram describing knot theory in terms of its corresponding
algebra, braid group and polynomial invariant. The paper also
identifies nontrivial unitary braid representations with  universal
quantum gates, and derives a Hamiltonian to determine the evolution
of a universal quantum gate, and further computes the Markov trace
in terms of a universal quantum gate for a link invariant to detect
linking numbers.

 }

\end{center}

\begin{tabbing}



\end{tabbing}

\newpage

\section{Introduction}

In a vector space $V$, there are rich algebraic structures over the
direct sum of its tensor products $V^{\otimes n}, n\in {\mathbb N}$,
for example, the braid relation describing knot theory
\cite{kauffman0}. In the vector space $V^{\otimes n}$, two types of
operations can be defined: either global or local. For example, the
transpose in $V^{\otimes n}$ is a global operator on $V^{\otimes n}$
itself, while the partial transpose is a local operator in
$V^{\otimes n}$ and acts on the subspace of $V^{\otimes n}$. The
paper focuses on the permutation $P$ and its partial transpose
$P_\ast$ and tries to exhaust their underlying algebraic and
topological properties.

The paper's goals take root in quantum information theory
\cite{nielsen}. The partial transpose itself has become a standard
tool in quantum entanglement theories for detecting the separability
of a given quantum state, see \cite{wolf} for more references. The
Peres--Horodecki criterion \cite{peres,horodecki} says that the
partial transpose of a separable density operator is positive.
 A state $\rho$ is called separable or ``classically correlated",
i.e., convex combinations of product density operators
 \cite{werner0, werner1}, \eq \rho=\sum_i
 \lambda_i \, \rho_A^{(i)} \otimes \rho_B^{(i)}, \,\, \Theta_2(\rho)=\sum_i
 \lambda_i \, \rho_A^{(i)} \otimes \Theta(\rho_B^{(i)}),\,\, \sum_i \lambda_i=1,
 \,\, \lambda_i \ge 0 \en
where the symbol $\Theta_2$ denotes the partial transpose,  the
symbol $\Theta$ denotes the transpose and $\rho_A$, $\rho_B$ are
states for subsystems $A$, $B$, respectively.

Here the Werner state \cite{werner1} is identified as a rational
solution of the Yang--Baxter equation (YBE) \cite{yang, baxter},
i.e., $Id + u P$ as the linear combination of identity and
permutation, while the isotropic state \cite{horodecki1} is found to
form a braid representation, i.e., $Id + v P_\ast$ with a specified
parameter $v$. Permutation as an element of the group algebra ${\cal
S}_n$ of the symmetric group $S_n$ and its partial transpose form a
new algebra called the $PPT_n$ algebra which is isomorphic to the
Brauer algebra \cite{brauer}. It plays important roles in
constructing multipartite symmetric states \cite{werner2, eggeling}
in quantum information theory. In terms of Brauer diagrams,
complicated computations can be simplified, for example, proving
that quantum data hiding is at least asymptotically secure in the
large system dimension \cite{eggeling, werner3}.

Recently, knot theory is involved in the study of quantum
information theory. A series of papers  explore natural similarities
 between topological entanglement and quantum entanglement, see
\cite{kauffman2, molin1, molin2} for universal quantum gates and
unitary solutions of the YBE; see \cite{kauffman3, kauffman5,
kauffman8} for quantum topology and quantum computation; see
\cite{kauffman4, kauffman7} for quantum entanglement and topological
entanglement; see \cite{kauffman9} for teleportation topology. They
identify nontrivial unitary solutions of the YBE with universal
quantum gates.

Now let $P$ be a swap permutation matrix specified by $P(|\xi\rangle
\otimes |\eta\rangle )=|\eta\rangle\otimes |\xi\rangle$ and let the
$\check{R}$-matrix be a unitary solution to the braid relation (the
braid version of the YBE). Examples are the following forms  \eq
\check{R} = \left(
\begin{array}{cccc} a & 0 & 0 & 0 \\ 0 & 0 & d & 0
\\ 0 & c & 0 & 0 \\ 0 & 0 & 0 & b \end{array} \right), \qquad
\tau =\check{R} P= \left( \begin{array}{cccc} a & 0 & 0 & 0 \\ 0 & c
& 0 & 0
\\ 0 & 0 & d & 0 \\ 0 & 0 & 0 & b \end{array} \right),
\en where $a, b, c, d$ can be any scalars on the unit circle in the
complex plane. From the point of view of braiding and algebra,
$\tau$ is a solution to the algebraists version of the YBE with
$\tau = \check{R} P$, and $P$ is to be regarded as an algebraic
permutation or as a representation of a virtual or flat crossing.
Then from the point of view of quantum gates, we have the phase gate
$\tau$ and the swap gate $P$. The $\check{R}$-matrix can be used to
make an invariant of knots and links that is sensitive to linking
numbers.

The virtual braid group \cite{kauffman10, kauffman11, kauffman12,
kamada} is an extension of the classical braid group by the
symmetric group. Each virtual braid operator can be interpreted as a
swap gate. With virtual operators in place, we can compose them with
the $\check{R}$-matrix to obtain phase gates and other apparatus in
quantum computation. Therefore the virtual braid group provides a
useful topological language for building patterns of quantum
computing.

Besides applications of permutation and its partial transpose to
quantum information theory, there are unexpected underlying
algebraic and topological structures. Here the permutation's partial
transpose $P_\ast$ is recognized as an idempotent of the
Temperley--Lieb ($TL$) algebra \cite{lieb}. The projectors in terms
of $Id$, $P$ and $P_\ast$ suggest the concept of the $D_n$ tangle
allowing both classical and virtual crossings which generalizes the
$T_n$ tangle only having classical crossings \cite{kauffman15}. It
is well known that braids can be represented in the $TL$ algebra
\cite{jones1, jones2, jones3}. The linear combinations of $Id$, $P$
and $P_\ast$ are found to form braid representations, for example
the isotropic state $Id + v P_\ast$; flat braid representations
underlying common solutions of the braid relation and YBEs; unitary
braid representation via Yang--Baxterization \cite{jones} and a
general unitary braid representation observed from a solution of the
coloured YBE \cite{murakami1, molin4}.

In view of a series of results in this paper, we articulate the
concept of the virtual Temperley--Lieb algebra which forms a virtual
braid representation similar to the $TL$ algebra representation of
the braid group. We define a generalized Temperley--Lieb algebra
which is isomorphic to the Brauer algebra in the graphical sense. As
a natural summary, we draw the ABPK diagram describing knot theory
in terms of its corresponding algebra, braid and polynomial to
emphasize roles of  the virtual Temperley--Lieb algebra in the
virtual knot theory.

 The plan of our paper is organized as follows.
 Section 2 interprets permutation's partial transpose as an
 idempotent of the $TL$ algebra and introduces the concept of the
 $D_n$ tangle. Section 3 observes the Werner state and
 isotropic state from the point of YBE solutions under dual
 symmetries. Section 4 defines the $PPT_n$ algebra and lists the
 axioms of the Brauer algebra with an example generated by the
 permutation $P^{\pm}$ and its partial transpose's deformation $Q_\ast$.
 Section 5 presents the family of virtual braid groups by sketching
 axioms defining virtual, welded and unrestricted braid groups. Section 6
 applies the linear combinations of $Id$, $P$ and $P_\ast$ to virtual
 braid representations and YBE solutions, and proposes the virtual
 Temperley--Lieb algebra and show it in the $ABPK$ diagram.
 Section 7 identifies nontrivial unitary braid representations
 with universal quantum gates and calculates the Markov trace
 for a link invariant to support such an identification.
 The last section concludes the paper and makes comments on
 further research. The appendix A sketches the Hecke algebra
 representation of the braid group. The appendix B provides a proof
 for Theorem 1.

\begin{figure}
\begin{center}
\epsfxsize=12.cm \epsffile{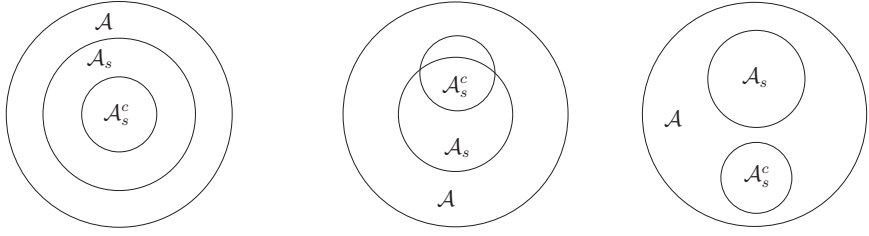} \caption{The commutant
${\cal A}^c_s$ of the subalgebra ${\cal A}_s$ of the algebra
 $\cal A$. } \label{fig0}
\end{center}
\end{figure}

 The braid representation $\sigma$-matrix and
 YBE solution $\check{R}$-matrix
 are $d^2 \times d^2$ matrices acting on $ V\otimes V$ where $V$ is
 an  $d$-dimensional complex vector space. The $\sigma$-matrix ($\check{R}$-matrix)
 is essentially a generalization of permutation \cite{yang, baxter}.
 The symbols $\sigma_i$ and $\check{R}_i$ denote $\sigma$ and $\check{R}$ acting
 on the tensor product $V_i\otimes V_{i+1}$. The symbols $Id$ or $1\!\! 1$
 denote the identity map from $V$ to $V$. The commutant of the algebra
 ${\cal A}_s$, a subalgebra of the algebra ${\cal A}$,  is the set
 ${\cal A}_s^c$  of elements of the algebra ${\cal A}$ commuting with
 all elements of the subalgebra ${\cal A}_s$, so that it is either a subset
 or independent of  ${\cal A}_s$ or intersects ${\cal A}_s$, see Figure 1.

\section{Permutation and its partial transpose }

We define the permutation $P$, and realize its partial transpose
$P_\ast$ to be an idempotent of the $TL$ algebra. We combine $Id$,
$P$ and $P_\ast$ into projectors, propose the concept of the $D_n$
tangle and finally make diagrammatic representations for the
extended Temperley--Lieb algebra.

\begin{figure}[!hbp]
\begin{center}
\epsfxsize=8.cm \epsffile{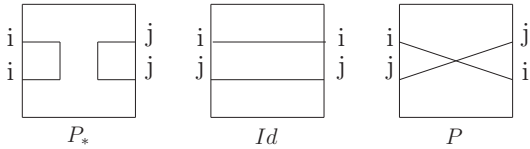} \caption{Permutation,
permutation's partial transpose and identity.} \label{fig1}
\end{center}
\end{figure}

\subsection{Permutation's partial transpose}

For two given independent finite Hilbert spaces ${\cal H}_1$ and
${\cal H}_2$ with bases $\{|i\rangle\}$ and $\{|j\rangle\}$
respectively, the tensor products $|i\rangle \otimes | j \rangle$
denoted by $|ij\rangle$, $ i,j=1,\cdots, d$, gives the product basis
for ${\cal H}_1 \otimes {\cal H}_2$. The permutation operator $P$
has the form as $P=\sum_{i,j=1}^d |i j \rangle \langle j i |$ which
satisfies $P |\xi \eta\rangle=|\eta\xi \rangle$. See Figure 2. The
partial transpose operator $\Theta_2$ is defined by acting on the
operator product $A\otimes B$ and only transforming indices
belonging to the basis of the second Hilbert space ${\cal H}_2$,
namely $\Theta_2(A\otimes B)=A\otimes B^T$. When the basis of ${\cal
H}_2$ is fixed, the symbol $B^T$ denotes the transpose of the matrix
$B$. With the partial transpose $\Theta_2$ acting on the permutation
$P$, we have a new operator $P_\ast$ given by \eq \label{past}
P_\ast=\Theta_2\circ P=\sum^d_{i, j=1} (|i \rangle\otimes \langle
j|) (|j\rangle \otimes \langle i |)^T=d |\Omega\rangle\langle
\Omega|, \qquad |\Omega\rangle=\frac 1 {\sqrt{d}} \sum_{i=1}^d |i i
\rangle \en where $|\Omega\rangle$ is called the Schmidt form and
$P_\ast$ acts on $|\xi\eta\rangle$ by \eq P_\ast |\xi
\eta\rangle=\sum_{i,j=1}^d
 |ii\rangle \langle j j |\xi \eta\rangle =
 \sum_{i=1}^d |ii\rangle \delta_{\xi \eta}.
 \en
\begin{figure}[!hbp]
\begin{center}
\epsfxsize=14.3cm \epsffile{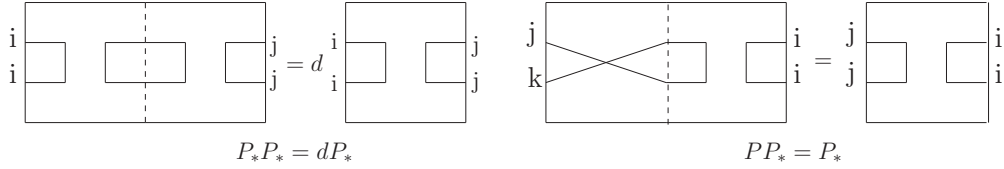} \caption{The products of
$P_\ast$ and $P$, $P_\ast$. } \label{fig2}
\end{center}
\end{figure}
 The permutation $P$ and its partial transpose $P_\ast$
 satisfy $P P_\ast=P_\ast P=P_\ast$ and $P_\ast P_\ast=d\,P_\ast$,
 see Figure 3. In the four dimensional case (d=2), the operators $P$
 and $P_\ast$ have the forms in matrix:
 \eq P=\left(\begin{array}{cccc}
 1 & 0 & 0 & 0 \\
 0 & 0 & 1 & 0  \\
 0 & 1  & 0 & 0 \\
 0 & 0 & 0 & 1
 \end{array}\right), \qquad P_\ast=\left(\begin{array}{cccc}
 1 & 0 & 0 & 1 \\
 0 & 0 & 0 & 0  \\
 0 & 0  & 0 & 0 \\
 1 & 0 & 0 & 1
 \end{array}\right).
\en
\begin{figure}
\begin{center}
\epsfxsize=12.5cm \epsffile{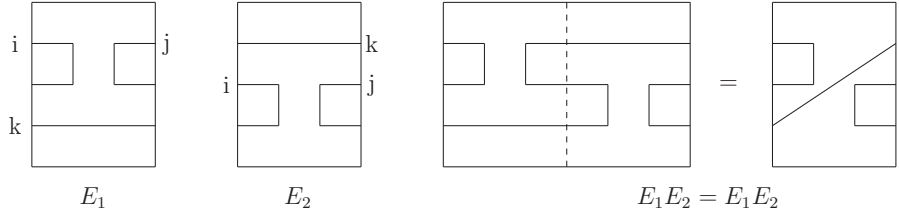} \caption{The generators $E_1$
and $E_2$ of the $TL_3$ algebra.} \label{fig3}
\end{center}
\end{figure}

The permutation's partial transpose $P_\ast$ is an idempotent of the
$TL$ algebra. The $TL_n(\chi)$ algebra is generated by $1\!\! 1$ and
$n-1$ hermitian operators $E_i$ satisfying \eqa \label{tl}
TLR(\chi):\,\,
 E_i^2 &=& \chi E_i, \qquad (E_i)^\dag=E_i,\,\,\, i=1,\ldots,n-1, \nonumber\\
 E_i E_{i\pm1} E_i &=& E_i, \qquad E_i E_j=E_j E_i, \,\,\, |i-j|>1,
  \ena
which is denoted as ``$TLR(\chi)$" for the Temperley--Lieb relation
with a loop parameter $\chi$. We check that $P_\ast$ satisfies the
axioms of the $TL_n(d)$ algebra,
  \eq P_\ast P_\ast=\sum^d_{i,j,i^\prime, j^\prime} |i
i\rangle \langle j j|i^\prime i^\prime \rangle \langle j^\prime
j^\prime |=d \sum_{i,j}^d |i i\rangle \langle j j|=d P_\ast. \en
Define the generators $E_1, E_2$ of the $TL_2(d)$ algebra in terms
of $P_\ast$, \eq E_1 =P_\ast \otimes Id, \qquad E_2=Id \otimes
P_\ast, \qquad E_1^2=d E_1, \,\, E_2^2=d E_2, \en see Figure 4.
After a little algebra, we have
 \eq
  E_1 E_2 E_1|ijk\rangle =\sum_{l=1}^d |llk\rangle \delta_{ij}=E_1
  |ijk\rangle,
 \en
similar to $E_2 E_1 E_2=E_2$.
\begin{figure}
\begin{center}
\epsfxsize=12.5cm \epsffile{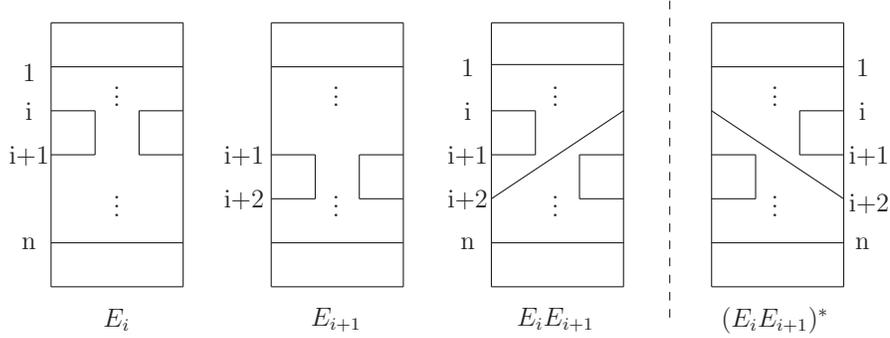} \caption{The generators
$E_i$ and $E_{i+1}$ of the $TL_n$ algebra.} \label{fig4}
\end{center}
\end{figure}
 Here are the generators of the $TL_n(d)$ algebra:
\eqa & & E_0=(Id)^{\otimes (n)}, \,\, E_1=P_\ast \otimes
(Id)^{\otimes (n-2)},\,\, E_2=Id\otimes P_\ast \otimes
(Id)^{\otimes (n-3)} \nonumber\\
 & & E_i= (Id)^{\otimes (i-1)} \otimes P_\ast \otimes (Id)^{\otimes (n-i-1)},
 \cdots,
 E_{n-1}=(Id)^{\otimes (n-2)} \otimes P_\ast.
 \ena
 See Figure 5 for their graphical representations.

\subsection{Projectors and $D_n$ tangles}

 We construct projectors in terms of
  $Id$, $P$ and $P_\ast$. The operators
 $\frac {P_\ast} d $, $\frac {1\!\! 1-P} 2$ and
 $\frac 1 2 (1\!\! 1+P)-\frac 1 d P_\ast$
 form a  set of projection operators, satisfying
 \eqa
  P_1 &=& \frac {P_\ast} d, \qquad P_2=\frac {1\!\! 1-P} 2, \qquad
 P_3=\frac 1 2 (1\!\! 1+P)-\frac 1 d P_\ast, \nonumber\\
 P_1^2 &=& P_1, \qquad P_2^2=P_2, \qquad P_3^2=P_3, \nonumber\\
  P_1 P_2 &=& 0, \qquad P_1 P_3=0, \qquad P_2 P_3=0, \nonumber\\
  & & P_1+P_2+P_3 = 1\!\! 1.
 \ena
The operators $\frac {P_\ast} d$ and $1\!\! 1- \frac {P_\ast} d$
also form a set of projectors,
 \eq (\frac {P_\ast} d)^2=\frac {P_\ast} d, \qquad
  (1\!\! 1- \frac {P_\ast} d)^2=1\!\! 1- \frac {P_\ast} d, \qquad
  \frac {P_\ast} d (1\!\! 1- \frac {P_\ast} d)=0.  \en
In addition, the operators $1\!\! 1-\frac 2 d P_\ast$
 and $P-\frac 2 d P_\ast$ are permutation-like,
 \eq (1\!\! 1-\frac 2 d P_\ast)^2=1\!\!
 1=P^2=(P-\frac 2 d P_\ast)^2. \en

We extend our construction of projectors in terms of $Id$, $P$ and
$P_\ast$ via the concept of the $D_n$ tangle. A $D_n$ tangle
contains classical or virtual crossings with $n$ begin-points in a
top row and $n$ end-points in a bottom row. Here the virtual
crossing refers to permutation. A $D_n$ tangle without any crossings
is also called an elementary tangle $T_n$ \cite{kauffman15}. A $D_n$
tangle with only virtual crossings is called a virtual tangle $P_n$
\cite{dye1}. We set up examples for the $T_2$ tangle, $P_2$ tangle
and $D_2$ tangle in Figure 6.
 \begin{figure}
 \begin{center}
 \epsfxsize=13.5cm \epsffile{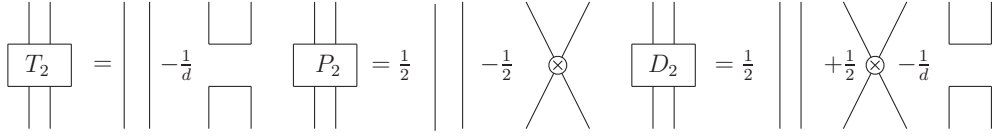} \caption{
 Projectors in terms of $2$ tangles.} \label{fig5}
 \end{center}
 \end{figure}
 Following a recursive procedure of deriving  the Jones--Wenzl projector $T_n$
 from a given $T_2$ diagram \cite{kauffman15}, we can generalize our construction
 $T_2$, $P_2$, $D_2$ to $T_n$, $P_n$, $D_n$, respectively.

 At $d=2$, the operators $1\!\! 1-P_\ast$ and $P-P_\ast$ are
 permutation-like. The operators $\frac {P_\ast} 2$,
  $\frac {1\!\! 1-P} 2$ and $\frac {1\!\! 1+P-P_\ast } 2$ provide a set of
 projectors and operators $\frac {P_\ast} 2$, $1\!\! 1- \frac
 {P_\ast} 2$ also. The projectors play basic roles in the
 Yang--Baxter theory, especially in six-vertex models
 \cite{kauffman0}. A solution of the
 YBE in terms of the set of projectors simplifies involved
 calculation and make related geometric (algebraic) descriptions clear.
 We apply projectors via $P$ and $P^\ast$ to study new quantum algebras
 in eight-vertex models \cite{molin1, molin2}. Our results will
 appear elsewhere \cite{yong1}.

\subsection{On the $TL_n$ algebra and $TL$ diagrams}

One of the best known representations of the $TL$ algebra, where, in
terms of diagrams, each cup and each cap is a Kronecker delta, has
been used in lots of physics literature (e.g. Wu and collaborators
on the intersecting string model for Potts type models \cite{wu1,
wu2}). This representation gives a series of specializations of the
Jones polynomial by making the variable in the bracket model fit the
dimension of the representation (We also do some of this calculation
in the present paper). Our work sets up a representation of the $TL$
algebra in terms of permutation's partial transpose, which has a
natural diagrammatic representation. It is worthwhile examining the
potentiality of the partial transpose for setting up braid
representations.

As a matter of fact, Figures 2--5 naturally exhibit $TL$ diagrams
\cite{kauffman15, kauffman14}. They assign diagrammatical
descriptions to the $TL_n$ algebra. Such a diagram is a planar
$(n,n)$ diagram including a rectangle in the plane with $2 n$
distinct points: $n$ on its left (top) edge and $n$ on its right
(bottom) edge which are connected by disjoint strings drawn within
in the rectangle. The identity is the diagram with all strings
horizontal (vertical), while $E_i$ has its $i$th and $i+1$th left
(top) (and right (bottom)) boundary points connected and all other
strings horizontal (vertical). The multiplication $E_i E_j$
identifies right (bottom) points of $E_i$ with corresponding left
(top) points of $E_j$, removes the common boundary and replaces each
obtained loop with a factor $\chi$. The adjoint of $E^\ast_i$ is an
image under mirror reflection of $E_i$ on a vertical (horizontal)
line. So we have horizonal (vertical) $TL$ diagrams for showing the
$TL$ algebra. Here, we take horizonal $TL$ diagrams for the
multiplication of elements of the $TL$ algebra and vertical $TL$
diagrams for explaining braids or crossings.

 \section{The YBE solutions under dual symmetries}

We sketch various formulations of the YBE and then study the Werner
state \cite{werner1}, and the isotropic state \cite{horodecki1} for
examples of solutions of the YBE under dual symmetries. The Werner
state has the form of $Id+ u P$, $u$ being the parameter and its
partial transpose is called the isotropic state $Id+ u P_\ast$.

\subsection{The YBEs and Yang--Baxterization}

 The YBE \cite{yang, baxter} was originally found in the procedure
 of achieving exact solutions of two-dimensional quantum field theories or
 lattice models in statistical physics. It has the form \eq \label{qybepl}
 \check{R}_i(x)\,\check{R}_{i+1}(xy)\,\check{R}_i(y)=
 \check{R}_{i+1}(y)\,\check{R}_i(xy)\,\check{R}_{i+1}(x)\en with the
 asymptotic condition $\check{R}_i (0)=\sigma_i $ and $x$ called the
 spectral parameter. In terms of  new parameters $u, v$ given by  $x=e^{u}$ and $y=e^v$,
 the YBE (\ref{qybepl}) has the other form \eq  \label{qybead}
 \check{R}_i(u)\,\check{R}_{i+1}(u+v)\,\check{R}_i(v)=
 \check{R}_{i+1}(v)\,\check{R}_i(u+v)\,\check{R}_{i+1}(u).
 \en
 Furthermore, the algebraic YBE mentioned in the introduction reads
 \eq
 R_{12}(x)R_{13}(x y)R_{23}(y)=R_{23}(y)R_{13}(x y)R_{12}(x), \en
 $R_{ij}$ acting on  $V_i \otimes V_j$, which has a solution by
 $R(x)=\check{R}(x)P$.

 Taking the limit of $x\to 0$ leads to the braid relation
 from the YBE (\ref{qybepl}) and the $\sigma$-matrix from the
 $\check{R}$-matrix.  Note that both $\sigma_i$ and
 $\check{R}_i(x)$ are fixed up to an overall scalar factor.
 Concerning relations between the $\sigma$-matrix and
 $x$-dependent solutions of the YBE (\ref{qybepl}), we construct
 the $\check{R}(x)$-matrix from a given $\sigma$-matrix. Such a construction is
 called Yang--Baxterization \cite{jones}. It is important to
 make distinctions between the braid relations and YBEs. The braid relations are
 topological but the YBE relations are not necessarily topological due
 to the spectral parameter.

\subsection{The Werner state and isotropic state}

 The Werner state $\rho_W$ \cite{werner1} for a biparticle has the form
 \eq \rho_W=p\, |\psi^{-}\rangle \langle \psi^{-}|+\frac {1-p} 4 {1\!\!
 1}_4 \en where the symbol ${1\!\! 1}_4$ is a four by four unit matrix and
 the Bell state $|\psi^{-}\rangle$ takes the  form
 $|\psi^{-}\rangle=\frac 1 {\sqrt{2}} (|01\rangle-|10\rangle)$.
So the Werner state $\rho_W$ has the form in matrix
  \eq \rho_W=
 \left(\begin{array}{cccc}
\frac {1-p} 4 & 0 & 0 & 0 \\
0    &  \frac {1+p} 4 & \frac {1-3 p} 4 & 0 \\
0  &  \frac {1-3 p} 4  & \frac {1+p} 4  & 0 \\
0   &  0 & 0 & \frac {1-p} 4
\end{array}\right).
 \en
Set $p=\frac {1-2 f} 3$, then the Werner state $\rho_W$ is a linear
combination of ${1\!\! 1}_4$ and $P$, namely, \eq \rho_W=\frac 1 6
((2-f) {1\!\! 1}_4+ (2 f-1)P) \en which has a generalized form in
$d$-dimension, \eq \rho_W=\frac 1 {d (d+1)} ((d-f) {1\!\! 1}_d+ (d\,
f-1)P). \en

Choosing $u=\frac {d f-1} {d-f}$, $f\neq d$, we obtain \eq \rho_W
(u)= \frac {d-1} {d (u+d)} ({1\!\! 1}_d+u\, P) \en which is a well
known rational solution $\check{R}(u)$-matrix of the YBE
(\ref{qybead}). For a given $\check{R}(u)$-matrix, a standard ``RTT"
relation procedure \cite{molin1, molin2} specifies a Hamiltonian
calculated by $H=\frac {d } {d u} {\check R} (u)|_{u=0}$.
 The Werner state $\rho_W$ is related to a Hamiltonian given by
 \eq H=\sum_{i=1}^N P_{i, i+1}=\frac 1 2 \sum_{i=1}^N
(1\!\! 1_4+ \sigma^x_i \otimes\sigma^x_{i+1}+ \sigma^y_i
\otimes\sigma^y_{i+1}+ \sigma^z_i \otimes\sigma^z_{i+1} ) \en
 which is the Hamiltonian of  the $XXX$ spin chain and where the Pauli
 matrices $\sigma^x$, $\sigma^y$ and $\sigma^z$ have the conventional
 formalism \eq \sigma^x=\left(\begin{array}{cc} 0 & 1
 \\ 1 & 0 \end{array} \right),\qquad
 \sigma^y =\left(\begin{array}{cc} 0 &
 -i \\ i & 0 \end{array} \right), \qquad \sigma^z
 =\left(\begin{array}{cc} 1 & 0 \\ 0 & -1 \end{array} \right). \en

The isotropic state $\rho_I$ is the partial transpose of the Werner
state $\rho_W$,  \eq \rho_I(v)=1\!\! 1_d + v P_\ast =\Theta_2
(\rho_W). \en It forms a braid representation when the parameter $v$
satisfies \eq \label{vb}
    v_\pm=-\frac 1 2 (d\mp \sqrt{d^2-4}), \qquad d\ge 2.
  \en
See the subsection 6.1 for the proof. The corresponding
$\check{R}_\pm(u)$-matrix via Yang--Baxterization \cite{kulish} has
the form \eq \check{R}_\pm (u) =u \rho_{I}(v_\pm) -u^{-1}
\rho_{I}^{-1}(v_\pm)=( u v_\pm -u^{-1} v_\mp ) + (u-u^{-1}) P_\ast,
\en which determines a local Hamiltonian with nearest neighbor
interactions, \eq H_{i,i+1}=\frac 1 2 1\!\! 1_4+\frac 1 2
(\sigma_i^x \otimes \sigma_{i+1}^x -\sigma_i^y \otimes
\sigma_{i+1}^y + \sigma_i^z \otimes \sigma_{i+1}^z). \en

\subsection{On the YBE solutions under dual symmetries}

Classical invariant theory tells us that the linear combinations $a
Id + b P$ of identity and permutation are the only operators
commuting with all unitary operators of the form $U\otimes U$.
Similarly, the operators $a Id + b P_*$ span the commutant of the
operators $U\otimes \overline U$, where $\overline{U}$ denotes the
complex conjugate of the matrix $U$ in the standard basis, which we
have fixed throughout. Moreover, the linear combinations of $Id,P$
and $P_*$ span the commutant of the operators $U\otimes U$ with $U$
real orthogonal. These facts are heavily used in the
characterization of symmetric states in quantum information theory,
where the states with the two kinds of symmetry are called Werner
states \cite{werner1} and isotropic states \cite{horodecki1},
respectively \cite{werner2}.

It is very natural to look at such symmetries for constructing
solutions of the braid and YBE relations. Indeed, we can require all
$\check{R}(u)$ to commute with $U\otimes U$ for all unitaries in an
appropriate subgroup $G$ of the unitary group.  This property has
then an immediate extension to relations on $n$ strands, and also
automatically admits the ordinary permutation operators to serve as
virtual generators. Obviously, the construction of solutions then
proceeds by first decomposing the representation $U\otimes U$ into
irreducible representations of $G$, typically leaving a much lower
dimensional space in which to solve the required non-linear
equations.

The choice of the operators $Id,P,P_*$ is by no means arbitrary, but
reflects the choice of orthogonal symmetry as the underlying
symmetry group $G$ for the single strand. Computations in the
commutant of the operators $U^{\otimes n}$ on $n$ strands are
simplified by choosing a basis, whose multiplication law can be
represented graphically \cite{eggeling}. As usual, we can write a
permutation $\pi$ as a (flat) braiding diagram, with $n$ strands
going in at positions $(1,\ldots,n)$ and coming out at
$(\pi_1,\ldots,\pi_n)$. This corresponds to the operator \eq
V_\pi=\sum_{i_1,\ldots,i_n}^d
           |i_{\pi_1}\cdots i_{\pi_n}\rangle\,
            \langle i_1,\cdots,i_n|.\en
If we apply a partial transposition to any tensor factor, a pair of
corresponding indices are swapped between the ket and the bra factor
of each term. Thus we are still left with a sum over $n$ indices,
each of which appears exactly twice, but the distribution of these
$2n$ indices over ket and bra is completely arbitrary. The
multiplication works exactly as for permutations, with the new
feature that strands can turn back. Also closed loops can appear,
which appear in the product only as a scalar factor $d$.

Note that such graphical representations are the Brauer diagrams
since the Brauer algebra $D_n$ \cite{brauer} maps surjectively to
the commutant of the action of the orthogonal group on the tensor
powers of its  representation. The Brauer diagrams are similar to
the $TL$ diagrams but allow strings to intersect. The next section
focuses on the Brauer algebra and the $PPT_n$ algebra generated by
permutation's partial transposes.

\section{The $PPT_n$ algebra and Brauer algebra}

In terms of $Id$, $P$ and $P_\ast$, we can set up multipartite
symmetric states under transformations of unitary group, orthogonal
group and the tensor product of unitary group and its complex
conjugation. Since the Peres--Horodecki criterion
\cite{peres,horodecki} involves a state and its partial transpose,
we figure out the set of permutation's partial transposes and
articulate it as the $PPT$ algebra. This algebra plus permutations
whose partial transposes are not themselves is isomorphic to the
Brauer algebra \cite{brauer}. Here we define the $PPT_n$ algebra,
explain it with examples, sketch the axioms of the Brauer algebra
and draw Brauer diagrams to show permutation and its partial
transpose.

\subsection{The $PPT_n$ algebra}

The transpose $\Theta$  in the $n$-fold finite dimensional Hilbert
space ${\cal H}^{\otimes n}$ is defined by its action on a given
operator
 \eq \Theta(|i_1 i_2 \cdots i_n \rangle
\langle j_1 j_2 \cdots j_n |)=|j_1 j_2 \cdots j_n \rangle \langle
i_1 i_2 \cdots i_n |, \en while the partial transpose $\Theta_k$,
$1\le k \le n$ takes the action, \eq \Theta_k(|i_1 \cdots i_k \cdots
i_n \rangle \langle j_1 \cdots j_k \cdots j_n |)=|j_1 \cdots j_k
\cdots j_n \rangle \langle i_1 \cdots i_k \cdots i_n |, \en so that
the multi-partial transpose $\Theta_{l_1 l_2\cdots\, l_j }$ is
defined as \eq \Theta_{l_1 l_2\cdots\, l_j
}=\Theta_{l_1}\Theta_{l_2}\cdots \Theta_{l_j}, \qquad
\Theta=\prod_{i=1}^n \Theta_k, \qquad 1\le j < n. \en Note that the
set of $\Theta_k$ forms an abelian group defined by
 $\Theta_k^2 =Id, \Theta_i \Theta_j =\Theta_j \Theta_i$.

The symmetric group $S_n$ consists of all possible permutations of
$n$-objects. Its group algebra ${\cal S}_n$ is generated by cyclic
permutations $\pi_i=(i,i+1)$ and $Id$,  satisfying \eqa
\label{symmetric}
   \pi_{i}  \pi_{i+1} \pi_{i} &=& \pi_{i+1} \pi_{i} \pi_{i+1}, \qquad
   i=1, \cdots, n-1, \nonumber\\
    \pi_{i}^2 &=& Id, \qquad \pi_{i} \pi_{j} = \pi_{j} \pi_{i},  \qquad
    j \neq i \pm 1.
 \ena
 Denote an operator $V_j$ in ${\cal H}^{\otimes n}$ in terms
 of Dirac's bras and kets,
 \eq
V_j=\sum_{i_1,\ldots,i_n}^d
           |i_1, \cdots, i_{j+1} i_j,\cdots i_n\rangle\,
            \langle i_1,\cdots, i_j i_{j+1}, \cdots, i_n|
            \en
which satisfies the following relations given by \eqa
   V_{i}  V_{i+1} V_{i} &=& V_{i+1} V_{i} V_{i+1}, \qquad
   i=1, \cdots, n-1, \nonumber\\
    V_{i}^2 &=& Id, \qquad V_{i} V_{j} = V_{j} V_{i},  \qquad
    j \neq i \pm 1
 \ena
 so that the set of $V_j$ forms a representation of the symmetry
 group algebra ${\cal S}_n$.

The action of $\Theta_j$ on $V_j$ leads to an idempotent $E_j$ given
by \eq E_j= \Theta_j (V_j)=\sum_{i_1,\ldots,i_n}^d
           |i_1, \cdots, i_{j} i_j,\cdots i_n\rangle\,
            \langle i_1,\cdots, i_{j+1} i_{j+1}, \cdots, i_n|
\en  satisfying the $TLR(d)$ relation with the loop parameter $d$
similar to the permutation's partial transpose $P_\ast$ presented
before. The $PPT_n(i)$ algebra denoting the action of $\Theta_i$ on
the ${\cal S}_n$ algebra have the generators as
  \eq
 V_1, V_2, \cdots, E_{i-1}, E_{i}, \cdots V_{n-1}.
\en For the multi-partial transpose $\Theta_{l_1 l_2\cdots\, l_j}$,
the $PPT_n(l_1 l_2\cdots l_j)$ algebra is generated in the same way
as the $PPT_n(i)$ algebra. The $PPT_n(i)$ algebra is isomorphic to
the symmetric group algebra ${\cal S}_n$ generated by $Id$ and $V_j$
since their generators are in one to one correspondence by
$V_{i-1}\to E_{i-1}$, $V_{i}\to
 E_i$. The $\Theta_k$ acting on the product $V_i V_j$ has
 the form
 \eqa
 \Theta_k (V_i^2) &=& Id, \qquad \Theta_{k+1} (V_k V_{k+1})=\Theta_{k+1} (V_{k+1})
 \Theta_{k+1} (V_{k})
  \nonumber\\
 \Theta_{k} (V_i V_{j}) &=& \Theta_{k} (V_i) \Theta_{k}
  (V_{j}),\qquad \textrm{for other cases}
 \ena
 which show that $\Theta_1$ is a homomorphism between the generators of
 ${\cal S}_n$ and those of $PPT_n$, the transpose $\Theta$ is an
 anti-homomorphism but the $\Theta_k, k \ge 2$ is neither homomorphism nor
 anti-homomorphism.

 \begin{figure}
 \begin{center}
 \epsfxsize=13.cm \epsffile{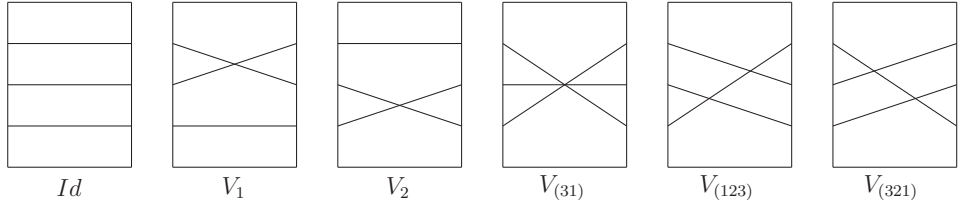} \caption{Brauer diagrams for
 the symmetric group algebra ${\cal S}_3$.} \label{fig6}
 \end{center}
 \end{figure}

The Brauer algebra is generated by $TL$ idempotents and virtual
crossings (i.e., permutations) and it can be also generated by a
$TL$ idempotent plus the symmetric group algebra ${\cal S}_n$. For
example, the idempotent $E_i$ is related to $E_{i-1}$ in the way \eq
\Theta_i(V_i)=E_i=V_{i-1} V_i E_{i-1} V_i V_{i-1}=V_{i-1} V_i
\Theta_{i} (V_{i-1}) V_i V_{i-1}. \en So the $PPT_n(i)$ algebra at
$n>2$ is a subalgebra of the Brauer algebra. The $PPT_n(i)$ algebra
 together with $V_{i-1}$ and $V_i$ denoted by the $\overline{PPT}_n(i)$ algebra
 is isomorphic to the Brauer algebra, similarly for
the algebra $\overline{PPT}_n(l_1 l_2\cdots l_j)$.

 \subsection{Examples for the $PPT_n$ algebra}

\begin{figure}
 \begin{center}
 \epsfxsize=13.cm \epsffile{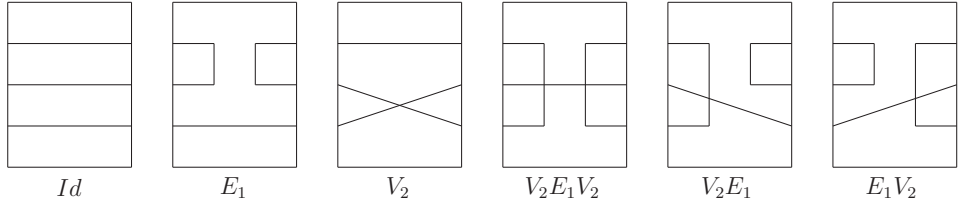} \caption{Brauer diagrams for
 the $PPT_3$ algebra.} \label{fig7}
 \end{center}
 \end{figure}

 The symmetric group $S_3$ is the set given by
 \eq
  S_3=\{e, (12), (23), (31), (123), (321)\}
 \en
including the identity $e$. It has two generators $\pi_1$ and
$\pi_2$ yielding the other elements by \eq (13)=\pi_2 \pi_1 \pi_2,
\,\, (123)=\pi_2 \pi_1, \,\, (321)=\pi_1 \pi_2. \en Introduce a
representation of ${\cal S}_3$ via a map \eq D: \pi \to
V_\pi=\sum_{i,j,k=1}^d |\pi(ijk)\rangle \langle ijk|, \qquad \pi\in
S_3 \en and every $V_\pi$ has its own Brauer diagram, see Figure 7.
 \begin{figure}
 \begin{center}
 \epsfxsize=9.cm \epsffile{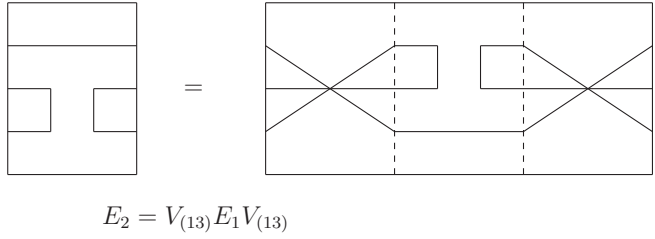} \caption{The transformation between
 two idempotents $E_1$ and $E_2$.} \label{fig8}
 \end{center}
 \end{figure}
The partial transpose $\Theta_1$ is a homomorphism between
 ${\cal S}_3$ and $PPT_3(1)$, see Figure 8, \eqa
 E_1 &=& \Theta_1 (V_{(12)}), \qquad V_2=\Theta_1 (V_{(23)})=V_{(23)}, \qquad
 \Theta_1 (V_{(13)})=V_2E_1V_2, \nonumber\\
 Id &=& \Theta(V_e), \qquad \Theta_1 (V_{(123)})=V_2E_1, \qquad \Theta_1
 (V_{(321)})=E_1 V_2
 \ena
where the generator $E_1$ is an idempotent and the other generator
$V_2$ is a permutation, \eq E_1^2= d E_1, \qquad V_2^2=1\!\! 1,
\qquad E_1V_2E_1=V_2. \en The $PPT_3(1)$ algebra, generated by $Id$,
$E_1$ and $V_2$, and $V_1$ form the $\overline{PPT}_3(1)$ algebra
which is isomorphic to the Brauer algebra $D_3(d)$.  For example,
the idempotent $E_2$ of $D_3(d)$ can be generated by $V_{(13)} E_1
V_{(13)}$ and $V_{(13)}=V_1 V_2 V_1$, see Figure 9. The $PPT_3(1)$
algebra minus the generator $V_2$ is isomorphic to the $TL_2(d)$
algebra.
 \begin{figure}
 \begin{center}
 \epsfxsize=13.cm \epsffile{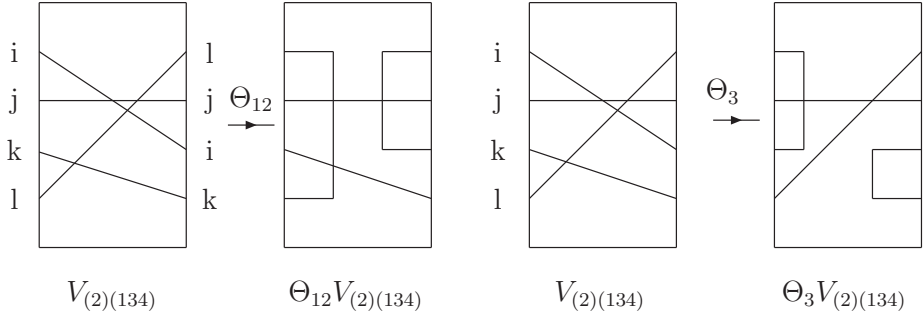} \caption{Brauer diagrams for
 partial transposes in the $PPT_4$ algebra.} \label{fig9}
 \end{center}
 \end{figure}

At $n=4$, consider the permutation element $(2)(134)$ of the
symmetric group $S_4$. The partial transpose $\Theta_{12}$ denotes
the transpose in the first and second Hilbert spaces. They have the
following presentations \eq
 V_{(2)(134)}=\sum_{i,j,k,l=1}^d |ljik\rangle \langle i j k l|, \qquad
 \Theta_{12} V_{(2)(134)}=\sum_{i,j,k,l=1}^d |ijik\rangle \langle
 ljkl|.
\en See Figure 10. In terms of Brauer diagrams, we calculate the
product $ V_{(2)(134)} \Theta_{12} V_{(2)(134)}$ and recognize it as
the adjoint of $\Theta_{3} V_{(2)(134)}$ where $\Theta_{3}$ denotes
the transpose in the third Hilbert space,
 \eq
 \Theta_{3} V_{(2)(134)}=\sum_{i,j,k,l=1}^d |ljkk\rangle \langle
 ijil|,
 \en
and we also show $\Theta_{12} V_{(2)(134)}$ to be an idempotent
which allows intersecting strings, see Figure 11.
\begin{figure}
 \begin{center}
 \epsfxsize=13.5cm \epsffile{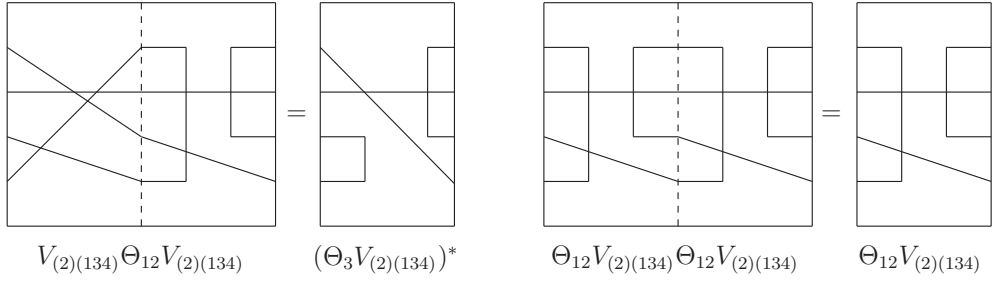} \caption{Brauer diagrams
 for multiplications in the $PPT_4$ algebra} \label{fig10}
 \end{center}
 \end{figure}

 \subsection{The axioms of the Brauer algebra}

 Here we list the axioms of the Brauer algebra $D_n(x)$ \cite{brauer}. The parameter $x$
 called the loop parameter takes the dimension $d$ in this
 paper. Its generators have $TL$ idempotents $E_i$ and virtual
 crossings $v_i$, $i=1,\cdots, n-1$, see Figure 12.
 \begin{figure}
\begin{center}
\epsfxsize=11.5cm \epsffile{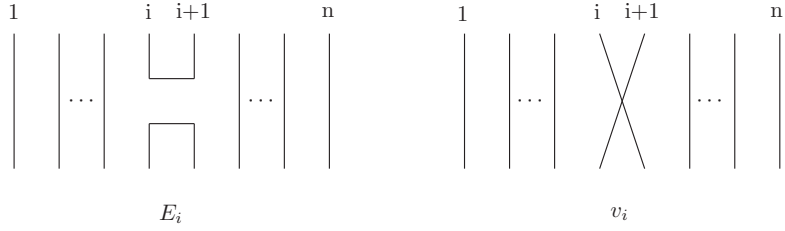}
 \caption{Generators $E_i$ and $v_i$ of the Brauer algebra.} \label{fig11}
\end{center}
\end{figure}
 $TL$ idempotents $E_i$ satisfy the $TLR(x)$ relation (\ref{tl})
 and virtual crossings $v_i$ are defined by the virtual crossing
 relation denoted by ``$VCR$",
 \eqa \label{vbgr1}
 VCR:\,\, v_i^2 &=& 1\!\! 1, \qquad  v_i v_{i+1} v_i = v_{i+1} v_i v_{i+1},
 \nonumber\\
 v_i v_j &=& v_j v_i, \qquad j \neq i \pm 1.
\ena  They satisfy the mixed relations by \eqa \label{brauer}
 & & (ev/ve): E_i v_i = v_i E_i =E_i, \qquad
  E_i v_j = v_j E_i, \qquad j\neq i \pm 1,
 \nonumber\\
 & &(vee): v_{i\pm1} E_{i} E_{i\pm1} = v_{i} E_{i\pm1},
 \qquad (eev): E_{i\pm1} E_i v_{i\pm1}=E_{i\pm 1} v_i,
 \ena
for example, see Figure 13 for the $(eev)$ axiom. {\em The Brauer
algebra generated by idempotents $E_i$ and virtual crossings $v_i$
are defined by the $TLR(x)$, $VCR$, $(ev/ve)$, $(vee)$ and $(eev)$
relations.} These defining axioms can drive the other mixed
relations. The axioms $(vee)$ and  $(eev)$ lead to the $(vee)$ and
$(evv)$ relations, respectively, \eqa && (vee)
 \Rightarrow (vve): v_{i\pm 1} v_i E_{i\pm 1}=E_i
 E_{i\pm1},\nonumber\\
 & & (eev) \Rightarrow E_i E_{i\pm1} E_i v_{i\pm1}=E_i E_{i\pm1}v_i
 \Rightarrow (evv): E_i v_{i\pm1}
 v_i=E_i E_{i\pm1}.
 \ena
 Identifying $(vve)$ with $(evv)$ leads to the $(vev)$ relation,
 \eq
(vve), (evv) \Rightarrow v_{i\pm1}v_i E_{i\pm1}=E_i v_{i\pm1}v_i
\Rightarrow (vev): v_{i\pm1}E_i v_{i\pm1}=v_i E_{i\pm1} v_i.
 \en
Furthermore, the relations $(vev)$ and $(vve)$ suggest the relation
$(eve)$,
 \eq (vev) \Rightarrow E_i v_{i\pm1} E_i=v_{i\pm1}v_i E_{i\pm1} E_i=E_i E_{i\pm1}E_i
 \Rightarrow (eve): E_i v_{i\pm1}E_i=E_i. \en

 \begin{figure}
\begin{center}
\epsfxsize=13.cm \epsffile{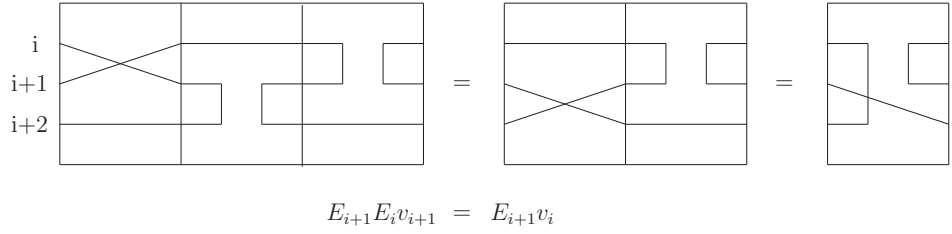}
 \caption{The $(eev)$ axiom of the Brauer algebra.} \label{fig12}
\end{center}
\end{figure}

At the diagrammatical level, it is explicit that the permutation $P$
and its partial transpose $P_\ast$ form the Brauer algebra.
Introduce a new permutation notation $P^\pm$ and denote $Q_\ast$ as
a $q$ deformation of permutation's partial transpose $P_\ast$. They
have the forms \eq \label{dppt} P^\pm=\left(\begin{array}{cccc} 1 &
0 & 0 &
 0 \\ 0 & 0 & \pm 1 & 0 \\
0 & \pm 1 & 0 & 0 \\
0 & 0 & 0 & 1
\end{array}\right), \qquad  Q_\ast=\left(\begin{array}{cccc} 1 & 0 & 0 &
 -q \\ 0 & 0 & 0 & 0 \\
0 & 0 & 0 & 0 \\
-q^{-1} & 0 & 0 & 1
\end{array}\right).
\en We check that $P^{\pm}$ and $Q_\ast$ form the Brauer algebra.
The $Q_\ast$ is a $TL$-idempotent satisfying \eq Q_\ast Q_\ast=2
Q_\ast, \qquad Q_{\ast i} Q_{\ast i\pm 1} Q_{\ast i} =Q_{\ast i}\en
and the hermitian $Q_\ast$ requires the deformed parameter $q$
living at the unit circle. The three mixed relations for the
defining axioms are verified by \eqa
 P^\pm Q_\ast &=& Q_\ast P^\pm = Q_\ast, \nonumber\\
P_i^{\pm} Q_{\ast i+1} Q_{\ast i} &=& P^{\pm}_{i+1} Q_{\ast
i},\qquad Q_{\ast i+1} Q_{\ast i} P^{\pm}_{i+1} =Q_{\ast i+1}
P^{\pm}_i. \ena

Note that in the following sections we will focus on $P_\ast$
instead of $Q_\ast$ since they behave in the same way. The Brauer
algebra is a limit of the Birman--Wenzl algebra
 \cite{wenzl2, murakami} in which the defining axioms and derived mixed
 relations are independent of each other.
The Birman--Wenzl algebra was devised to explain the Kauffman two
variable polynomial in terms of its trace functional.

\begin{figure}
\begin{center}
\epsfxsize=10.5cm \epsffile{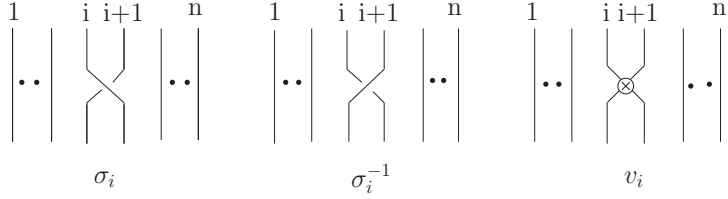} \caption{The braid generators
and virtual braid generator. } \label{fig13}
\end{center}
\end{figure}

\section{The virtual, welded and unrestricted braid groups }

Besides the commutant of the tensor power of the orthogonal group,
$PPT_n$ algebra and Brauer algebra, there are the virtual knot
theory and its underlying virtual $TL$ algebra via permutation and
its partial transpose. We sketch the axioms defining the family of
virtual braids, then construct the virtual braid representations in
terms of the linear combination of $Id$, $P$ and $P_\ast$.

The classical braid group $B_{n}$ (the Artin braid
 group) on $n$ strands is generated by the braids $\sigma_i$
 and it consists of all words of the form
 $ \sigma_{j_1} ^{ \pm 1} \sigma_{j_2} ^{\pm 1} ... \sigma_{j_n} ^{ \pm 1} $
 modulo the braid relations, see Figures 14-15: \eqa \label{bgr}
  BGR:\,\, \sigma_{i}  \sigma_{i+1} \sigma_{i} &=& \sigma_{i+1} \sigma_{i} \sigma_{i+1},
  \qquad i=1, \cdots, n-1, \nonumber\\
   \sigma_{i} \sigma_{j} &=& \sigma_{j} \sigma_{i}, \qquad  j \neq i \pm 1,
 \ena
which is denoted as ``$BGR$" for the braid group relation.
\begin{figure}
\begin{center}
\epsfxsize=11.5cm \epsffile{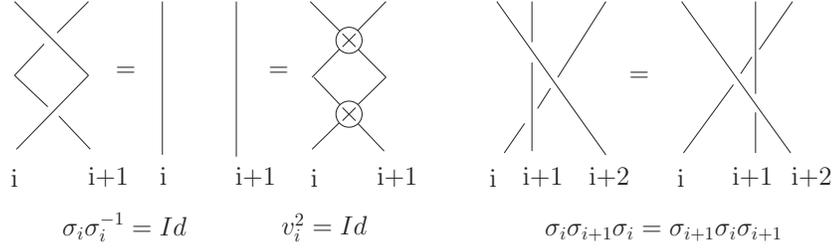} \caption{Identity
and the braid relation.} \label{fig14}
\end{center}
\end{figure}
The virtual braid group $VB_n$ \cite{kauffman10, kauffman11,
kauffman12, kamada} is an extension of the classical braid group
$B_n$ by the symmetric group $S_n$. It has both the braids
$\sigma_i$ and virtual crossings $v_i$. A virtual crossing $v_i$ is
represented by two crossing arcs with a small circle placed around
the crossing point. In virtual crossings, we do not distinguish
between under and over crossing but which are described respectively
in the classical knot theory.
\begin{figure}
\begin{center}
\epsfxsize=12.5cm \epsffile{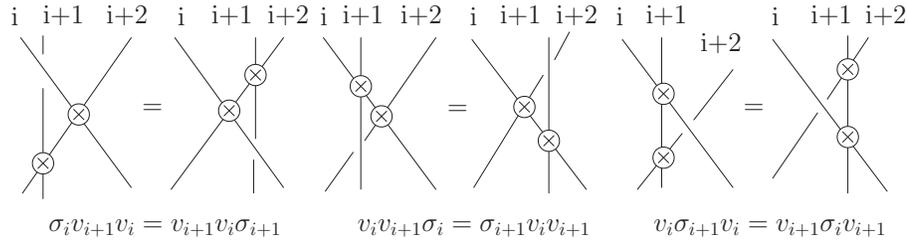} \caption{The
virtual braid relations. } \label{fig15}
\end{center}
\end{figure}
The virtual generators $v_i$ satisfy the $VCR$ relation
(\ref{vbgr1}) and they form a representation of the symmetric group
$S_n$. The virtual generators $v_i$ and braid generators $\sigma_j$
satisfy the mixed relations: \eqa \label{vbgr2}
 VBR: \,\,  \sigma_i v_j &=& v_j \sigma_i, \qquad   j \neq i \pm 1,
 \nonumber\\ v_i \sigma_{i+1} v_i &=& v_{i+1} \sigma_i v_{i+1},
\ena which is denoted as ``$VBR$" for the virtual braid relation.
Here the second mixed relation is also called the special detour
relation.

There are the following  relations also called special detour moves
for virtual braids and they are easy consequences of (\ref{vbgr1})
and (\ref{vbgr2}), see Figure 16: \eqa \label{relation}
\sigma_{i}^{\pm}v_{i+1}v_{i} &=& v_{i+1}v_{i} \sigma_{i+1}^{\pm}, \nonumber\\
v_{i}v_{i+1} \sigma_{i}^{\pm} &=& \sigma_{i+1}^{\pm}v_{i}v_{i+1}, \nonumber\\
v_{i} \sigma_{i+1}^{\pm}v_{i} &=& v_{i+1} \sigma_{i}^{\pm}v_{i+1}.
\ena This set of relations taken together defines the basic
isotopies for virtual braids.
\begin{figure}
\begin{center}
\epsfxsize=11.5cm \epsffile{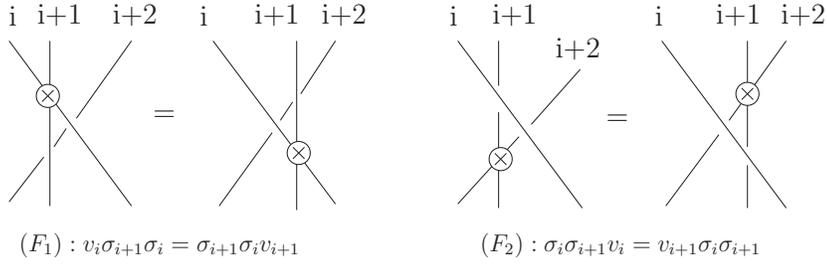} \caption{The forbidden
moves: $(F_1)$ and $(F_2)$. } \label{fig16}
\end{center}
\end{figure}
The move with two real crossings and one virtual crossing is a
forbidden move in the virtual knot theory. However, there are two
types of forbidden moves: the one with an over arc denoted by
$(F_1)$ and the other with an under arc denoted by $(F_2)$, \eq
\label{forbidden} (F_1): v_{i} \sigma_{i+1} \sigma_{i} =
\sigma_{i+1} \sigma_i v_{i+1}, \qquad  (F_2): \sigma_i \sigma_{i+1}
v_{i} = v_{i+1} \sigma_{i} \sigma_{i+1}, \en see Figure 17. The
first forbidden move $(F_1)$ preserves the combinatorial fundamental
group, as is not true for the second forbidden move $(F_2)$. This
makes it possible to take an important quotient of the virtual braid
group $VB_n$. The welded braid group $WB_n$ on $n$ strands
\cite{FRR} satisfies the same isotopy relations as the $VB_n$ group
but allows the forbidden move $(F_1)$. The unrestricted virtual
braid group $UB_n$ allows the forbidden moves $(F_1)$ and $(F_2)$
although any classical knot can be unknotted in the virtual category
if we allow both forbidden moves \cite{KANENOBU, NELSON}.
Nevertheless, linking phenomena still remain.

 The shadow of some link in three dimensional space without specifying
 the weaving of that link is a link or knot diagram which does not distinguish
 the over crossing from the under one. The shadow crossing without regard to
 the types of crossing is called a flat crossing. The flat virtual braid group
 $FV_n$ \cite{kauffman11} consists of virtual crossings
 $v_{i}$ and flat crossings $c_{i}$. It satisfies the same relations as
 the $VB_n$ except the
 braid $\sigma_i$ replaced with the flat crossing $c_i$ satisfying $c_i^2=1\!\! 1$.
The generalization of the $FV_n$ is called the flat unrestricted
braid group $FU_{n}$. It is the quotient of the $FV_{n}$ by adding
the forbidden move of $FV_{n}$.  Note that for the $FV_{n}$ there is
only one type of forbidden move since here $(F_1)$ is the same as
$(F_2)$, see, \eq
 (F_1) \Rightarrow \sigma_{i+1} v_i \sigma_{i+1} = \sigma_i v_{i+1} \sigma_i
 \Rightarrow (F_2).
\en

\begin{figure}
 \begin{center}
 \epsfxsize=8.5cm \epsffile{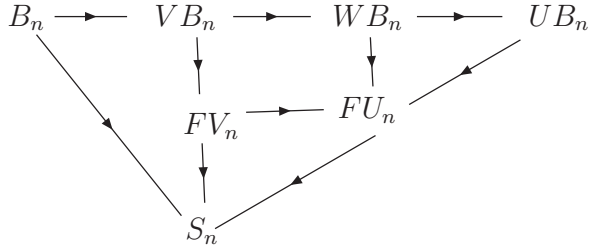} \caption{Relationships among
 various braid groups.} \label{fig17}
 \end{center}
 \end{figure}

The flat unrestricted braid group $FU_{n}$ is a quotient of the
welded braid group $WB_{n}$, obtained by setting all the squares of
the braiding generators equal to $1\!\! 1$. Thus there is a
surjective homomorphism from $WB_{n}$  to $FU_{n}.$ This
homomorphism is a direct analogue of the standard homomorphism from
the braid group $B_n$ to the symmetric group $S_n$. Figure 18 draws
a commutative diagram of these relationships where all structures
map eventually to the symmetric group $S_{n}$ \cite{kauffman13}.

\section{The virtual braid representations via $P$ and $P_\ast$}

We set up virtual braid representations in terms of $Id$, $P$ and
$P_\ast$. The isotropic state $Id + v_\pm P_\ast$ satisfies the
braid relation. The braid (virtual crossing) $1\!\! 1-P_\ast$ and
virtual crossing (braid) $P$ form flat unrestricted braid
representations which underlie common solutions of the braid
relation and YBEs. The linear combination of $1\!\! 1-P_\ast$ and
$P$ leads to a family of unitary braid representations with
adjustable parameters.

\subsection{The braid representation via permutation's partial transpose}

 Denote the operator $1\!\! 1+v P_\ast, v\neq 0$ by
$\check{R}(v)$. Substituting $\check{R}(v)$ into both sides of the
braid relation (\ref{bgr}), we obtain \eqa & &
\check{R}_i(v)\check{R}_{i+1}(v)\,
  \check{R}_{i}(v)|ijk\rangle
  = |ijk\rangle +(2 v + v^3+d v^2) \sum_{l=1}^d |llk\rangle
 \delta_{ij}  \nonumber\\
 & & + v \sum_{l=1}^d |ill\rangle\delta_{jk}
 +v^2 \sum^d_{l=1}|k l l\rangle
   \delta_{ij} + v^2 \sum_{l=1}^d |lli\rangle\delta_{jk},
   \nonumber\\
 & &\check{R}_{i+1}(v)\,\check{R}_i(v)\,
 \check{R}_{i+1}(v)|i j k\rangle
=|i j k\rangle+ (2 v+d v^2 + v^3) \sum_{l=1}^d
  |ill\rangle \delta_{jk} \nonumber\\
& & +v\sum_{l=1}^d |llk\rangle \delta_{ij} + v^2 \sum_{l=1}^d
|lli\rangle \delta_{jk} + v^2\sum_{l=1}^d |kll\rangle \delta_{ij}.
 \ena
Identifying both sides leads to an equation of the variable $v$, \eq
2 v+d v^2 + v^3=v \Rightarrow v+v^{-1}=-d  \en which has solutions
$v_\pm$ given by (\ref{vb}). So $1\!\! 1+v_\pm P_\ast$ or
$v_\mp\,1\!\! 1+P_\ast$ forms a family of braid representations. The
$P_\ast$ itself does not form a braid representation since the terms
with the coefficient $v^3$ on both sides are different. In addition,
the operator $\check{R}(v)$ is not a solution of the YBE
(\ref{qybepl}) or (\ref{qybead}). But $1\!\! 1+u P, u\neq 0$ is a
rational solution of the YBE (\ref{qybepl}) but not a solution of
the braid relation. Note the calculation
 \eq (\check{R}(v))^2 =1\!\! 1+
 v (2+ v d) P_\ast= \check{R}(v) + v(1+ v d) P_\ast.
 \en
It says: at $v=-\frac 1 d$, the $\check{R}(v)$ is a projector; at
$v=-\frac 2 d$, the $\check{R}(v)$ is permutation-like, and
represents a braid only for $d=2$.

Note that a solution of the braid relation can be constructed in
terms of a $TL$ idempotent,  see the appendix A for the detail. Here
we have $\check R_\pm=v_\mp 1\!\! 1 + P_\ast$ so that the Hecke
condition \cite{kulish}  is satisfied by \eq
(\check{R}_\pm)^2=(v_\pm-v_\mp){\check R_\pm}+ 1\!\! 1, \qquad
 v_\pm + ({v_\pm})^{-1}=-d.
\en
Note that for $d=3$ the orthogonal group coincides with the spin-$1$
representation of $SU(2)$. The basic technique of using strand
symmetry can of course also be extended to higher spin
representations of $SU(2)$. For example, at spin $3/2$, we get a
four dimensional commutant, in which the braid relation can be
solved.

\subsection{Flat unrestricted braid representations}

In \cite{kauffman10}, remarks about quantum link invariants show
that any braid group representation defined in the usual way by a
solution to the braid version of the YBE (ie. a solution to the YBE
that satisfies the braid relation) extends to a representation of
the virtual braid group when the virtual generator is represented by
the permutation (swap gate). Therefore the braid $1\!\! 1+ v_\pm
P_\ast$ and the virtual crossing $P$ form a representation of the
$VB_2$ group.

We construct a flat unrestricted braid representation $FU_2$. At
$d=2$ and $u=-1$, the operator $1\!\! 1-P_\ast$ denoted by $P^\ast$
forms a braid representation,
 \eq  P^\ast=1\!\! 1-\,
 P_{\ast}=\left(\begin{array}{cccc}
 0 & 0 & 0 & -1 \\
 0 & 1 & 0 & 0  \\
 0 & 0  & 1 & 0 \\
 -1 & 0 & 0 & 0
 \end{array}\right), \qquad P^\ast P^\ast=1\!\! 1.
 \en  and $P$
 The $P_\ast$ acting on a state $|\xi\eta\rangle$
 lead to
 \eq P |\xi\eta\rangle=|\eta\xi\rangle, \qquad
  P_\ast |\xi\eta\rangle=\sum_{l=0}^1 |ll\rangle
  \delta_{\xi\eta}, \qquad \xi, \eta=0, 1,
 \en
 so that the action of $P^{\ast}$ on $|ij\rangle$ takes the form \eq
 P^{\ast} |ij\rangle =|ij\rangle- (|00\rangle+|11\rangle)
 \delta_{ij}. \en
 In terms of $P$ and $P^\ast$, we verify the following equalities:
 \eq \label{fu21}
 P_i P_{i+1} P_i = P_{i+1} P_{i} P_{i+1}, \qquad P^\ast_i
 P^\ast_{i+1} P^\ast_i = P^\ast_{i+1} P^\ast_{i} P^\ast_{i+1}, \en
 which proves that $P$ and $P^\ast$ form braid representations; \eqa
 \label{fu22} P_i^2 &=& (P^\ast_i)^2= 1\!\! 1, \qquad    P_i P^\ast_{i+1} P_i =
 P_{i+1} P^\ast_i P_{i+1}, \nonumber\\
 P_i P^\ast_{i+1} P^\ast_{i} &=& P^\ast_{i+1} P^\ast_{i}
P_{i+1}, \qquad
  P^\ast_{i} P^\ast_{i+1} P_i = P_{i+1} P^\ast_{i} P^\ast_{i+1},
 \ena
 which proves that the flat crossing $P^\ast$ and virtual crossing $P$
 form a flat unrestricted braid representation
 $FU_2$; \eqa  \label{fu23} {P^\ast_i}^2 &=& P_i^2=1\!\! 1, \qquad
 P^\ast_{i} P_{i+1} P^\ast_{i} = P^\ast_{i+1}
 P_{ i} P^\ast_{i+1}, \nonumber\\
 P^\ast_{i} P_{i+1} P_i &=& P_{i+1} P_i P^\ast_{i+1}, \qquad
  P_i P_{i+1} P^\ast_{i} = P^\ast_{i+1} P_i P_{i+1},
 \ena
 which proves the flat crossing $P$ and virtual crossing $P^\ast$
 form the other flat unrestricted braid
 representation $FU_2$. Note that in higher dimensional ($d>2$) cases,
 substitute the braid $1\!\! 1+ v_\pm P_\ast$ and virtual crossing $P$
 into two forbidden moves (\ref{forbidden}), we find that they are not
 satisfied at $d>2$. Also, the $1\!\! 1+ v_\pm P_\ast$ can not set up a
 flat braid representation for $d>2$.

\subsection{Common solutions of (\ref{qybepl}),
(\ref{qybead}) and (\ref{bgr})}

We consider the linear combination of $Id$, $P$ and $P_\ast$ by \eq
\label{lpa} \check{R}(u) = a 1\!\! 1+u P+b P_\ast. \en Theorem 1
(below) presents a family of common solutions of the braid relation
(\ref{bgr}) and YBEs (\ref{qybepl}) and (\ref{qybead}). See Figure
19.

\begin{figure}[!hbp]
\begin{center}
\epsfxsize=13.5cm \epsffile{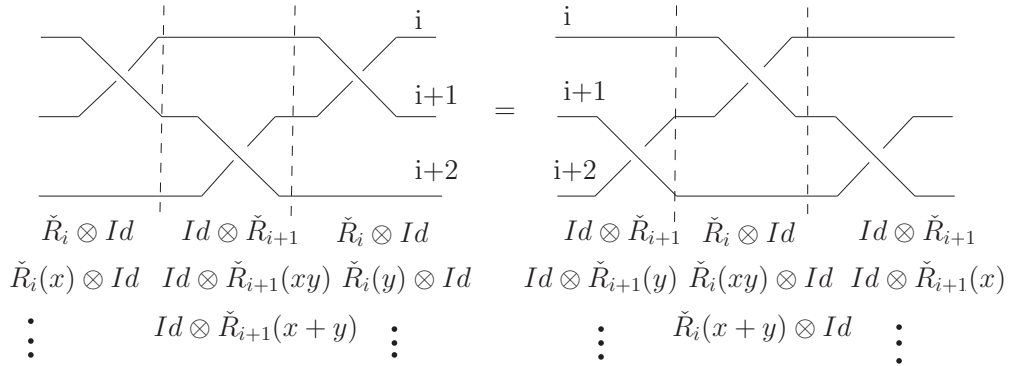} \caption{A common solution
of the braid relation and YBEs. } \label{fig18}
\end{center}
\end{figure}
{\bf Theorem 1.} The  $\check{R}(u)$ operator (\ref{lpa}) forms
 a braid representation (\ref{bgr}) only for $d=2,
  a=-b$ or $a=b=-u$ at $a\neq 0$ and it satisfies the
 YBE (\ref{qybepl}) or (\ref{qybead}) only for $d=2, a=-b$ at $a\neq 0$, the
 coefficient of  $P$ as the spectral parameter.
  \bigskip

Before proving Theorem 1, we make the statements of Theorem 1 clear.
The coefficient of $P_\ast$ being the spectral parameter leads to
trivial results. For $a=b=-u$ and $d=2$, the $\check{R}(u)$-matrix
 has the form \eq \check{R}=1\!\! 1-P+P_\ast
  =\left(\begin{array}{cccc}
  1 & 0 & 0 & 1 \\
  0 & 1  & -1 & 0 \\
  0 & -1  & 1  & 0 \\
  1 &  0 & 0 & 1  \\
 \end{array}\right)
 \en
 which is a known symmetric solution of the eight-vertex model.
For $d=2, a=-b$, the $\check{R}(u)$-matrix takes the form \eq
\label{matrixru}
 \check{R}(u) = a P^\ast+u P
 = \left(\begin{array}{cccc}
u & 0 & 0 & -a \\
0 & a & u & 0  \\
0 & u  & a & 0 \\
-a & 0 & 0 & u
 \end{array}\right),
\en which is a symmetric eight-vertex model with a parameter $u$ and
also satisfies the YBEs (\ref{qybepl}) and (\ref{qybead}), $u$ as
the spectral parameter.

Note that the braid $\check{R}(u)$-matrix (\ref{matrixru}) and the
virtual crossing $P$ form a unrestricted braid representation of
$UB_2$. They satisfy both forbidden moves (\ref{forbidden}) but the
$\check{R}(u)$-matrix has its square by \eq
 \check{R}(u)^2=(a^2+u^2)+2 a u (P+P^\ast).
\en  which is proportional to $Id$ only at $a u=0$.

Now we present the proof for Theorem 1. \bigskip

{\bf Proof}. The proof has two parts.  The first indicates that in
higher ($d>2$) dimension the $\check{R}(u)$ operator (\ref{lpa})
leads to trivial conclusions. We have to calculate all matrix
entries of both sides of (\ref{bgr}), (\ref{qybepl}) and
(\ref{qybead}) and then compare them one by one, see the appendix B
for details.

The second verifies that at $d=2$ the $\check{R}(u)$-matrix
(\ref{matrixru}) forms  common solutions of (\ref{bgr}),
(\ref{qybepl}) and (\ref{qybead}). The proof uncovers that flat
unrestricted braid representations generated by (\ref{fu21}),
(\ref{fu22}), (\ref{fu23}) underlie the existence of common
solutions of the braid relation (\ref{bgr}) and YBEs (\ref{qybepl}),
(\ref{qybead}). After a little algebra, we have \eqa
 \label{pl}
 & & \check{R}_i (x) \check{R}_{i+1} (z) \check{R}_i (y)=
  a^3 P_i^\ast P^\ast_{i+1} P^\ast_i + x y z  P_i P_{i+1} P_i
  \nonumber\\
  & +& a^2 y P^\ast_i P^\ast_{i+1} P_i + a^2 z P_i^\ast P_{i+1}
  P_i^\ast + a^2 x P_i P^\ast_{i+1} P^\ast_i
  \nonumber\\
 & + & a y z P^\ast_i P_{i+1} P_i +  a x y P_i P_{i+1}^\ast P_i + a x
 z P_i P_{i+1} P^\ast_i, \nonumber\\
 & &  \check{R}_{i+1}(y) \check{R}_i (z) \check{R}_{i+1} (x)
  = a^3 P^\ast_{i+1} P^\ast_i P^\ast_{i+1} + x y z P_{i+1} P_i
  P _{i+1} \nonumber\\
 &+& a^2 x P^\ast_{i+1} P^\ast_i P_{i+1} + a^2 z P^\ast_{i+1}
 P_i P^\ast_{i+1} + a^2 y P_{i+1} P^\ast_i P^\ast_{i+1} \nonumber\\
  &+& a x z P^\ast_{i+1} P_i P_{i+1} + a x y P_{i+1} P^\ast_i
  P_{i+1} +a y z P_{i+1} P_i P^\ast_{i+1}
 \ena
which shows that the $\check{R}(u)$-matrix satisfies the following
equation \eq \label{color}
 \check{R}_i (x) \check{R}_{i+1} (z) \check{R}_i (y)=
 \check{R}_{i+1}(y) \check{R}_i (z) \check{R}_{i+1} (x).
\en We finish the proof by replacing the symbol $z$ with $x y$ or
$x+y$ and requiring $x=y=z$. In addition, the equation (\ref{color})
is related to the coloured Yang--Baxter equation \cite{murakami1}.

$\hfill \Box $

\subsection{Unitary braid representations via Yang--Baxterization}

There is a bigger family of common solutions satisfying (\ref{bgr}),
(\ref{qybepl}) and (\ref{qybead}). We derive it via
Yang--Baxterization \cite{jones}. The $\check{R}(u)$-matrix
(\ref{matrixru}) divided by a scaling
 factor $a$ has the form, which is an  example of a general
 $\check{R}_\pm$-matrix by  \eq   \label{rpm}
 \check{R} =\left(\begin{array}{cccc}
  t & 0 & 0 & -1 \\
  0 & 1  & t & 0 \\
  0 & t  & 1  & 0 \\
  -1 &  0 & 0 & t  \\
 \end{array}\right), \qquad t=\frac u a, \qquad
 \check{R}_\pm =\left(\begin{array}{cccc}
  t & 0 & 0 & q \\
  0 & 1  & \pm t & 0 \\
  0 & \pm t  & 1  & 0 \\
  q^{-1} &  0 & 0 & t  \\
 \end{array}\right),
 \en $q$ being the deformation parameter.
 The $\check{R}_\pm$-matrix has three distinguished eigenvalues:
\eq \lambda_1=1+t, \,\,\, \lambda_2=1-t, \,\,\, \lambda_3=t-1. \en
Via Yang--Baxterization, the corresponding $\check{R}_\pm(x)$-matrix
is obtained to be \eqa \label{baxter}
 \check{R}_\pm(x) &=& \check{R}_\pm+ x (1-t^2) \check{R}_\pm^{-1} \nonumber\\
  &=& \left(\begin{array}{cccc}
  t (1-x) & 0 & 0 &  q (1+x) \\
  0 & 1+x  & \pm t (1-x) & 0 \\
  0 & \pm t (1-x)  & 1+x  & 0 \\
  q^{-1} (1+x) &  0 & 0 & t (1-x)  \\
 \end{array}\right).
 \ena The unitarity condition, with the normalization factor denoted by
 $\rho$,
 \eq \check{R}_\pm(x)\check{R}_{\pm}^\dag(\bar x)=\check{R}_{\pm}^\dag(\bar x)
  \check{R}_{\pm}(x)=\rho 1\!\!1,  \en
leads to $\|q\|^2=1$ and $\|x\|^2=1$ for real $t$, the symbol
$||\cdot||$ denotes the norm of a given complex number (function),
see \cite{molin1, molin2, molin3} for the detail of
Yang--Baxterization. We present Theorem 2 as follows.
   \bigskip

{\bf Theorem 2}. The $\check{R}_\pm(x)$-matrix (\ref{baxter})
 satisfies the braid relation (\ref{bgr}) and YBEs
 (\ref{qybepl}), (\ref{qybead}), $x$ as
 the spectral parameter. The virtual crossing $P$ and the braid $\check{R}_+(x)$
 (\ref{baxter}) form a unrestricted braid representation $UB_2$, while the
 virtual crossing $P$ and the braid $\check{R}_-(x)$ (\ref{baxter})
 form a virtual braid representation $VB_2$.

 \bigskip

 Theorem 2 is proved similar as Theorem 1.
 The permutation $P^\pm$ and a new permutation-like
matrix given by $Q^\ast=1\!\! 1-Q_\ast$,  satisfy flat braid
relations, \eqa \label{qfu21}
 P_i^\pm P_i^\pm &=& 1\!\! 1,  \qquad P^\pm_i P^\pm_{i+1} P^\pm_i
 = P^\pm_{i+1} P^\pm_{i} P^\pm_{i+1},
 \nonumber\\ Q_i^\ast Q_i^\ast &=& 1\!\! 1, \qquad Q^\ast_i Q^\ast_{i+1} Q^\ast_i
 = Q^\ast_{i+1} Q^\ast_{i} Q^\ast_{i+1}. \ena
The flat crossing $Q^\ast$ and virtual crossing $P^\pm$ form a flat
unrestricted braid representation
 $FU_2$:\eq
 \label{qfu22}  P^\pm_i Q^\ast_{i+1} P^\pm_i =
 P^\pm_{i+1} Q^\ast_i P^\pm_{i+1}, \qquad
 P^\pm_i Q^\ast_{i+1} Q^\ast_{i} = Q^\ast_{i+1} Q^\ast_{i}
 P^\pm_{i+1}.
 \en
The flat crossing $P^\pm$ and virtual crossing $Q^\ast$
 form the other flat unrestricted braid
 representation $FU_2$:
 \eq  \label{qfu23}
 Q^\ast_{i} P^\pm_{i+1} Q^\ast_{i} = Q^\ast_{i+1} P^\pm_{ i} Q^\ast_{i+1}, \qquad
 Q^\ast_{i} P^\pm_{i+1} P^\pm_i = P^\pm_{i+1} P^\pm_i
 Q^\ast_{i+1}.
 \en
In terms of $P^\pm$ and $Q^\ast$, the $\check{R}_\pm(x)$-matrix
(\ref{baxter}) is written as \eq \check{R}_\pm(x)= a(x) Q^\ast+ c(x)
P^\pm, \qquad a(x)=1+x,\,\, c(x)=t (1-x). \en Therefore the proof
for Theorem 2 also underlies unrestricted braid representations
specified by (\ref{qfu21}), (\ref{qfu22}) and (\ref{qfu23}).

 Furthermore, we introduce the coloured YBE
 \cite{murakami1, molin4} by \eq \check{R}_{i+1} (\mu, \nu) \check{R}_i (\lambda, \nu)
\check{R}_{i+1} (\lambda, \mu) = \check{R}_{i+1} (\lambda,\mu)
 \check{R}_i (\lambda,\nu) \check{R}_{i+1}(\mu,\nu).\en
Choose $\check{R}(\lambda,\mu)=\lambda P^\pm+\mu Q^\ast$. It
satisfies the coloured YBE because of unrestricted braid
representations generated by $P^\pm$ and $Q^\ast$. In terms of
$Q^\ast$ and $P^\pm$, the most general $\check{R}_\pm(X,Y)$-matrix
has the form $\check{R}_\pm(X,Y)=a(X) Q^\ast + c(Y) P^\pm$, $X, Y$
denoting involved parameters.  It is a solution of the braid
relation (\ref{bgr}). The unitary braid representation condition \eq
\check{R}_\pm^\dag\check{R}_\pm=(||a(X)||^2+||c(Y)||^2)+(\bar a (X)
c(Y)+\bar c (Y) a(X) ) (P^\pm -Q_\ast) \en requires $\bar a (X)
c(Y)+\bar c (Y) a(X)=0$ and the hermitian $Q^\ast$ leads to
$||q||=1$. For example, setting $a(X)=1+x$ and $c(X)=t(1-x)$, we
have \eq (t+\bar t)(1-||x||^2)+(t-\bar t)(\bar x-x)=0 \en which
derives $||x||=1$ for $t=\bar t$, consistent with (\ref{baxter}).

 \subsection{On unitary representations of the $TL_n$ and $B_n$}

 Via a family of unitary braid representations as above,
 we use it to compute knot invariants depending on
 adjustable parameters in order to detect connections between
 topological entanglements and quantum entanglements.
  Some representations of the $TL$ algebra \cite{kauffman3}
 are found to have interesting unitary representations and we
 explain how these are related to quantum computing and
 the Jones polynomial. They are an elementary construction for more
 general representations due to H. Wenzl \cite{wenzl3, wenzl4}.
 We now know a lot about what happens when one tries to make
 braid representations unitary. Unitary solutions
 to the braid relation (YBE without spectral parameter) are classified \cite{dye}.
 The upshot is that there are very few solutions that have any power
 for doing knot theory, but this is just for the standard representation.

\subsection{On the virtual Temperley--Lieb algebra}

  In terms of the permutation's partial transpose $P_\ast$, we set up
  a representation of the $TL_n$ algebra and the braid representation
  $1\!\! 1+ v_\pm P_\ast$. Choosing the virtual crossing $P$ and the braid
  $1\!\! 1+ v_\pm P_\ast$, we have a virtual braid representation.
  Underlying what we have done  is the algebra of $P$ and $P_\ast$.
  Regarding $P$ as a virtual crossing and $P_\ast$ as a $TL$-idempotent, we touch
  the concept of the virtual Temperley--Lieb ($vTL$) algebra.
  Thus it is of interest to articulate the $vTL$ algebra and axiomatize
  it in a relatively obvious way. This axiomatization is useful for
  understanding the extension of the Witten--Reshetikhin--Turaev invariant
  to the virtual knot theory \cite{dye1}.

  The virtual $TL$ algebra is an algebra underlying the
  virtual knot theory. From the graphical point, it
  is an algebra of all possible connections between $n$ points and
  $n$ points and is generated by the usual $TL$ generators plus an operator that behaves
  like a permutation operator and is diagrammed by two flat crossing strands.
  The $vTL$ algebra so obtained is surjective to the so-called Brauer
  algebra discovered by Brauer in the 1930's for the purpose
  of explicating invariants of the orthogonal group \cite{brauer},
  \cite{wenzl1}. Brauer had a diagrammatic for his algebra that is equivalent to the
 one we would get by extending the $TL$ algebra with virtual
 crossings, see the subsection 2.3. But to this day $TL$ diagrams
 (Kauffman diagrams \cite{kauffman16}) and Brauer diagrams are
 regarded as separate subjects for the most part. Diagrams
 representing the $vTL$ algebra is called $vTL$ diagrams. Similar to
 horizontal (vertical) $TL$ diagrams, there are also horizontal
 (vertical) $vTL$ diagrams. Different from horizontal (vertical) $TL$
 diagrams, horizontal (vertical) $vTL$ diagrams allows intersections
 of horizontal (vertical) lines.

  We recall historical developments of knot theory
  since Jones's original work in 1985. He constructed the braid representation
  via the Jones algebra or Temperley--Lieb algebra. In \cite{homfly} the
  HOMFLY polynomial was found for the braid representation via the two-parameter
  Hecke algebra and on the heels of \cite{homfly} the Kauffman
  two-variable polynomial \cite{kauffman17} was proposed.
  Birman and Wenzl \cite{wenzl2, murakami} generalized the skein relations of
  the Kauffman two variable polynomial  to an algebra generalizing the Hecke algebra
  (called now the Birman--Wenzl algebra $BW_n$) and this maps to the Brauer
  algebra \cite{brauer} in analogy to the map of the Hecke algebra to the group algebra
  of the symmetric group. Afterwards, the virtual knot theory was
  articulated by involving the symmetric group $S_n$
  \cite{kauffman10, kauffman11, kauffman12, kamada}, where the
  virtual generalizations of knot polynomial \cite{kauffman10, dye2, manturov} appeared.
  Here we draw a picture called the ``ABPK" diagram
  describing knot invariants in terms of related algebra,
  braid group and polynomial invariant, see Figure 20.
  The horizontal axis denotes ``Algebra",
  ``Braid" and ``Polynomial" while the vertical-axis denotes
  different  presentations of ``Knot".

\begin{figure}
\begin{center}
\epsfxsize=11.5cm \epsffile{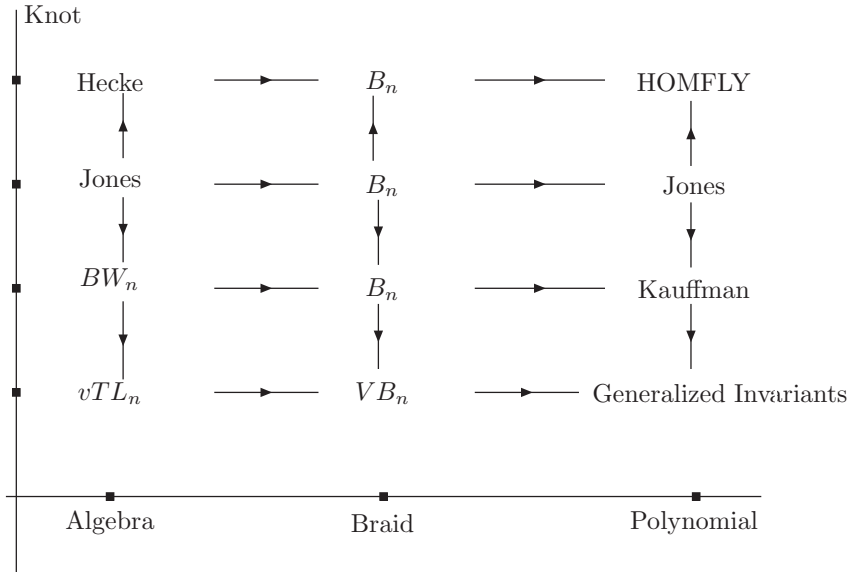}
 \caption{An $ABPK$ diagram showing knot theory} \label{fig20}
\end{center}
\end{figure}

\section{Universal quantum gate and unitary braid representation}

In terms of identity $Id$, the permutation $P^\pm$ and its deformed
partial transpose $Q_\ast$, we determine a family of unitary braid
representation. We recognize these as universal quantum gates and
write down the related Schr{\"o}dinger equation and with it
calculate the Markov trace for a link invariant to detect linking
numbers.

 \subsection{Universal quantum gate}

 A two-qubit gate $G$ is a unitary linear mapping  from  $V \otimes
 V$ to  $V \otimes V$ where $V$ is a two complex dimensional vector space. A gate
$G$ is said to be entangling if there is a vector $$| \alpha \beta
\rangle = | \alpha \rangle \otimes | \beta \rangle \in V \otimes V$$
such that $G | \alpha \beta \rangle$ is not decomposable as a tensor
product of two qubits.  The Brylinskis prove that a two-qubit gate
$G$ is universal iff it is entangling \cite{BB}.
 A pure state $|\psi\rangle$ is separable when
 \eq
|\psi\rangle=\sum^{1}_{i,j=0}a_{ij}|ij\rangle, \qquad
|ij\rangle=|i\rangle \otimes |j\rangle, \qquad a_{00} a_{11}\neq
a_{01} a_{10}.
 \en
The unitary $\check{R}(x)$-matrix acting on the state $|\psi\rangle$
has the form \eq
 \check{R} |\psi_{pt}\rangle =\sum^{1}_{i,j=0}\sum_{k,l=0}^{1}
 \check{R}^{kl}_{ij} a_{ij} |kl\rangle=
 \sum^{1}_{k,l=0}\,b_{kl}|kl\rangle.
\en  If there exists $a_{ij}$ leading to $b_{00} b_{11}=b_{01}
b_{10}$, then such the $\check{R}$-matrix can be recognized as a
universal quantum gate.

With a new variable $u$, the $\check{R}_\pm(u)$-matrix
(\ref{baxter}) has a simpler form \eq \label{gate}\check{R}_\pm(u)
=\left(\begin{array}{cccc}  u & 0 & 0 & q  \\
 0 & 1 & \pm  u & 0 \\
 0 &  \pm  u & 1 & 0 \\
 q^{-1}  & 0 & 0 &  u
\end{array}\right), \qquad  u=t\frac {1-x} {1+x}\en
which is a unitary matrix for $||q||^2=1$ and real $t$, $||x||^2=1$,
the latter two leading to imaginary $u$, i.e., $u=-\bar u$.
 It determines the coefficients $b^\pm_{ij}$ to be \eq
\left(\begin{array}{c} b^\pm_{00} \\ b^{\pm}_{01} \\ b^{\pm}_{10}
\\ b^\pm_{11}
\end{array}\right)= \left(\begin{array}{c}
  u \,\, a_{00}+q \,\, a_{11} \\  a_{01}\,\pm  u\, a_{10}
 \\  \pm  u\,\, a_{01} +a_{10}
\\ q^{-1} a_{00} +  u a_{11}
\end{array}\right)
\en and involved products given by
 \eqa
 b^\pm_{00} b^\pm_{11} &=& (1+ u^2)\, a_{00} a_{11}+ u  (q a^2_{11}+ q^{-1}
 a^2_{00}),
 \nonumber\\
 b^\pm_{01} b^\pm_{10} &=& (1+ u^2)\, a_{00} a_{11} \pm u
 (a^2_{01}+a^2_{10}).
 \ena
 As $u\neq0$, i.e., $x\neq 1$ and $t\neq 0$, the unitary $\check{R}(x)$-matrix
 (\ref{baxter}) is identified with a universal quantum gate.

\subsection{The Hamiltonian and unitary evolution}

Before deriving the Hamiltonian, we introduce the algebra of the
Pauli matrices. Denote two linear combinations of $\sigma_x$ and
$\sigma_y$ respectively by $\sigma_{n_1}$ and $\sigma_{n_2}$, \eq
\sigma_{n_1}= \cos{\frac \varphi 2}\,\, \sigma^x + \sin{\frac
\varphi 2} \,\, \sigma^y, \qquad \sigma_{n_2}=\cos{\frac
{\varphi+\pi} 2} \,\,\sigma^x + \sin{\frac {\varphi+\pi} 2}
\,\,\sigma^y  \en which have the corresponding tensor products, \eq
\sigma_{n_1} \otimes \sigma_{n_1}=
\left(\begin{array}{cccc} 0 & 0 &  0 & q \\
 0& 0 & 1 & 0 \\
 0 & 1 & 0 & 0 \\
 q^{-1} & 0 & 0 & 0 \\
 \end{array}\right),\,\,\,\,\,\,
\sigma_{n_2} \otimes \sigma_{n_2}
 =-\left(\begin{array}{cccc} 0 & 0 &  0 & q \\
 0& 0 & -1 & 0 \\
 0 & -1 & 0 & 0 \\
  q^{-1} & 0 & 0 & 0 \\
 \end{array}\right).
 \en
They satisfy the following formulas given by
 \eqa
 \sigma_{n_1} \sigma_{n_2} &=& i \sigma^z =-\sigma_{n_2} \sigma_{n_1} ,
 \qquad \sigma_{n_1} \sigma_{n_2}\otimes  \sigma_{n_1} \sigma_{n_2}
  = -\sigma^z \otimes \sigma^z, \nonumber\\
 \sigma_{n_1} \otimes \sigma_{n_1} &=& - \sigma_{n_2} \sigma^{z}\otimes
  \sigma_{n_2} \sigma^{z}, \qquad  \sigma_{n_2} \otimes \sigma_{n_2}
  = -\sigma_{n_1} \sigma^{z}\otimes  \sigma_{n_1} \sigma^{z}.
\ena

Here the $\check{R}(x)$-matrix (\ref{baxter}) involves the
normalization factor $\rho$. Choose $t=1$, then $\rho=4$. The
$\check{R}_+(x)$-matrix (\ref{baxter}) has the form of the tensor
product of the Pauli matrices, \eqa \check{R}_+(x)
 &=& \frac 1 2 1\!\! 1_4 -\frac 1 2 x \sigma^z \otimes \sigma^z
  +\frac 1 2 \sigma_{n_1} \otimes  \sigma_{n_1}
  - \frac 1 2 x \sigma_{n_2}
   \otimes  \sigma_{n_2} \nonumber\\
&=& 1\!\! 1_4- (\frac 1 2 -\frac 1 2 \sigma_{n_1} \otimes
\sigma_{n_1}) - x \sigma^z \otimes \sigma^z (\frac 1 2 -\frac 1 2
\sigma_{n_1} \otimes
\sigma_{n_1}) \nonumber\\
 &=& 1\!\! 1_4- H_+  - x  \sigma^z \otimes \sigma^z  H_+
 \ena
and similarly the $\check{R}_-(x)$-matrix (\ref{baxter}) has the
form \eq \check{R}_-(x) = 1\!\! 1_4- H_- -x \sigma^z \otimes
\sigma^z H_-
 \en
where the symbols $H_+$ and $H_-$ are given by \eq H_+=\frac 1 2
(1\!\! 1_4- \sigma_{n_1} \otimes \sigma_{n_1}), \qquad H_-=\frac 1 2
(1\!\! 1_4+ \sigma_{n_2} \otimes \sigma_{n_2}).
 \en

Considering three projectors $H_\pm$ and $P_z$ which satisfy \eqa &
& H_\pm^2=H_\pm,  \qquad P_z^2=P_z,\,\,
P_z =\frac 1 2(1\!\! 1_4+\sigma^z\otimes \sigma^z), \nonumber\\
 && P_z H_\pm =H_\pm P_z, \qquad  (P_z H_\pm)^2=P_z H_\pm,
\ena we represent the  $\check{R}_\pm(x)$-matrix (\ref{baxter}) by a
unitary exponential function
 \eqa
 \check{R}_\pm (\theta)
  &=& 1\!\! 1_4-H_\pm - e^{-i\theta} \sigma^z \otimes \sigma^z H_\pm \nonumber\\
   &=& (1\!\! 1_4-H_\pm+e^{-i\theta} H_\pm) (1\!\! 1_4- P_z H_\pm -P_z H_\pm) \nonumber\\
   &=& e^{-i \theta H_\pm} e^{-i\pi P_z H_\pm}
    = e^{-i (\theta+ \pi P_z) H_\pm}
\ena where a formula for the projector $A$ has been exploited, \eqa
 & & e^{i\alpha A}=\sum_{n=0}^\infty \frac {(i \alpha)^n} {n!} A^n
 = 1\!\! 1_4+ \sum_{n=1}^\infty \frac {(i \alpha)^n} {n!}A \nonumber\\
 & & =1\!\! 1_4-A+\sum_{n=0}^\infty \frac {(i \alpha)^n} {n!} A=1\!\! 1_4-A+e^{i\alpha}
 A. \ena

Let us derive the Hamiltonian to determine the unitary evolution of
a unitary quantum gate. Denote the state $\psi$ independent of the
time variable $\theta$. Its time evolution $\psi(\theta)$ is
specified by the $\check{R}(\theta)$-matrix (\ref{baxter}),
 $\psi(\theta)=\check{R}(\theta) \psi$ which leads to the
Schr{\"o}dinger equation, \eq i\,\frac {\partial \psi(\theta)}
{\partial \theta}=H(\theta) \psi(\theta), \qquad H(\theta)=i \frac
{\partial\check{R}(\theta)}{\partial \theta}
\check{R}^{\dag}(\theta).  \en Hence the Hamiltonian $H_\pm(\theta)$
 is the projector $H_\pm$ given before. The time evolution operator
 $U_\pm(\theta)$ has the form \eq U_\pm(\theta)=e^{-i H_\pm \theta } =1\!\! 1_4-
H_\pm + e^{-i\theta} H_\pm \en which can set up the CNOT gate with
the help of local unitary transformations or single qubit
transformations \cite{molin1}.

\subsection{The Markov trace as a link invariant}

A basic point of this paper is to recognize nontrivial unitary braid
representations as universal quantum gates. When a unitary braid
representation can detect a link or knot in topological context, it
often also has the power of entangling quantum states. Here we have
the unitary $\check{R}_\pm(u)$-matrix (\ref{gate}) which is a
universal quantum gate at $u\neq 0$. In the following, we calculate
the Markov trace which is a link invariant in terms of the
$\check{R}_\pm(u)$-matrix in order to show that at $u\neq 0$ we are
able to detect linking numbers.

\begin{figure}
\begin{center}
\epsfxsize=10.cm \epsffile{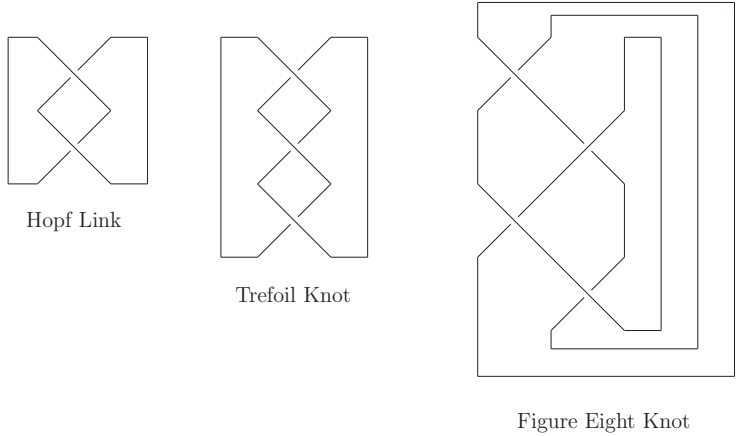} \caption{Examples for
representing links or knots by the closures of braids  }
\label{fig21}
\end{center}
\end{figure}

For a given link $L$, the link invariant for the Markov trace has
the form \eq Z(L)=\alpha^{- w(b)} Z_n (b), \qquad b\in B_n, \qquad L
\sim \bar{b}. \en The equivalence relation $L \sim \bar{b}$ says
that the link $L$ is isotopic to the closure of a braid $b$, as we
are told  by the Alexander theorem. For example, the Hopf link, the
Trefoil and the Figure Eight knot are represented by the closures of
the braids $\sigma_1^2$, $\sigma_1^3$ and $\sigma_1 \sigma_2^{-1}
\sigma_1 \sigma_2^{-1}$, respectively, see Figure 21. The writhe
$w(b)$ of the braid $b$ is the sum of the signs of crossings of the
braid $b$. Each under crossing $\sigma_i$ and over crossing
$\sigma_i^{-1}$ contribute $1$ and $-1$ to $w(b)$ respectively. For
example, the Hopf link, the Trefoil knot and the Figure Eight knot
have the writhe numbers of $2,3,0$, respectively. The normalization
factor $\alpha$ is determined by the specific choice of $Z_n(b)$
which is well defined on the braid group $B_n$ and satisfies the
following conditions
 \eqa
 Z_n(gbg^{-1}) &=& Z_n (b), \qquad g,b \in B_n, \nonumber\\
 Z_n (b\sigma^{\pm 1}_n) &=& \alpha^{\pm 1} Z_n (b),
 \qquad \sigma_n \in B_{n+1},
 \ena
the second equation also called the Markov move.

Now we set up the Markov trace in terms of the
$\check{R}_\pm(u)$-matrix (\ref{gate}). To avoid notation
ambiguities in this subsection, we denote the
$\check{R}_+(u)$-matrix by the $\check{R}$-matrix but in fact the
$\check{R}_-(u)$-matrix leads to the same link invariant. For the
generators $\sigma_i$ of the braid group $B_n$, the representation
$\rho_n(\sigma_i)$ has the form
 \eq
 \rho_n(\sigma_i)=Id^{\otimes i-1}\otimes \check{R}
 \otimes Id^{\otimes (n-i-1)}, \qquad i=1,\cdots n-1
 \en
and thus the Markov trace $Z(L)$ is chosen to be \eq
 Z(L)=\alpha^{- w(b)} Tr(\rho_n(b)).
\en The normalized factor $\alpha$ is calculated by the partial
trace $Tr_2 (\check{R})$ of the $\check{R}$-matrix, \eq
 Tr_2 (\check{R})= \alpha 1\!\! 1_2, \qquad   Tr_2 (\check{R})= \alpha^{-1} 1\!\!
 1_2.
\en
 The $\check{R}$-matrix has the form
 \eq
\check{R}=\left(\begin{array}{cccc} \check{R}^{00}_{00} &
\check{R}^{00}_{01} & \check{R}^{00}_{10}&
\check{R}^{00}_{11} \\[2mm]
\check{R}^{01}_{00} & \check{R}^{01}_{01} & \check{R}^{01}_{10}&
\check{R}^{01}_{11}
\\[2mm]
\check{R}^{10}_{00} & \check{R}^{10}_{01} & \check{R}^{10}_{10}&
\check{R}^{10}_{11}
\\[2mm]
\check{R}^{11}_{00} & \check{R}^{11}_{01} & \check{R}^{11}_{10}&
\check{R}^{11}_{11}
\end{array}\right),
 \en
and its partial trace $Tr_2 (\check{R})$ is given by \eq Tr_2
(\check{R}^{ai}_{bj})=\sum_{c=1}^2 {\check{R}^{ac}_{bc}} =
 \left(\begin{array}{cc} \check{R}^{00}_{00} + \check{R}^{01}_{01} &
\check{R}^{00}_{10}+ \check{R}^{01}_{11} \\
\check{R}^{10}_{00}+ \check{R}^{11}_{01} & \check{R}^{10}_{10}+
\check{R}^{11}_{11}
\end{array}\right).
\en If a reader is interested in the detail of the Alexander theorem
and the Markov theorem, please consult \cite{kauffman0} and
\cite{kauffman2}.

Before computing link invariants, we go through  the algebra of
$P^\pm$, $Q_\ast$ given by (\ref{dppt}) and $Q^\ast=1\!\!
1_4-Q_\ast$. They have the properties \eqa
 & & (Q^\ast)^{2 n}=(P^\pm)^{2n}=1 \!\! 1, \qquad P^\pm Q_\ast=Q_\ast P^\pm =Q_\ast,
 \nonumber\\
 & & P^\pm Q^\ast=P^\pm (1\!\! 1_4-Q_\ast)=P^\pm -Q_\ast=Q^\ast P^\pm,
  \qquad n\in {\mathbb N}
\ena
 along with the traces and partial traces of matrices,
\eqa
 & & Tr(P^\pm)=Tr(Q_\ast)=2, \qquad Tr(P^\pm Q_\ast)=0, \nonumber\\
 &&  Tr_2(P^\pm)=Tr_2(Q)_\ast=Tr_2(Q^\ast)=1.
\ena With the help of them, we represent the
$\check{R}_\pm(u)$-matrix (\ref{gate}) in terms of $P^\pm$ and
$Q^\ast$, and derive its inverse given by
 \eq \check{R}_\pm(u)=u P^\pm + Q^\ast, \qquad
\check{R}_\pm^{-1}(u)=\frac 1 {1-u^2} (-u P^\pm +Q^\ast) \en which
satisfy \eq
 \check{R}_\pm(u) + (1-u^2) \check{R}_\pm^{-1}(u)=2 (1\!\!
 1_4-Q_\ast),
\en
 leading to another normal way of computing a link invariant via the skein
 relation \cite{kauffman0}. The
 normalization factor $\alpha$ can be fixed by
 \eq
 Tr_2(\check{R}_\pm(u))=1+u, \qquad
 Tr_2(\check{R}^{-1}_\pm(u))=\frac 1 {1+u},\qquad \alpha=1+u
\en

As examples, we calculate the Markov traces corresponding to the
closures of the braids $\sigma_1^{2 n}$, $\sigma_1^{-2 n}$,
$\sigma_1^{2 n+1}$ and  $\sigma_1^{-2 n-1}$ with the writhe number
$2n$, $-2n$, $2n+1$ and $-2n-1$, respectively. It is well known that
$\overline{\sigma_1^{2n}}$ and $\overline{\sigma_1^{-2n}}$ are links
of two components with the linking numbers $n$ and $-n$. The linking
number denotes the half sum of the signs of crossings between two
components of a link. Also, $\overline{\sigma_1^{2n+1}}$ and
$\overline{\sigma_1^{-2n-1}}$ are knots for positive number $n$ and
they are unknots at $n=0$. The following trace formulas are helpful
in calculation, \eqa Tr(\check{R}_\pm^{2n})
  & =&  \sum_{k=0}^{2n} C_{2 n}^k u^k Tr ( (P^\pm)^k (Q^\ast)^{2n-k}
 ) \nonumber\\
 & &= 4 \sum_{l=0}^n C_{2n}^{2l} u^{2 l}=2 ((1+u)^{2n}+(1-u)^{2n}),
 \nonumber\\
Tr(\check{R}_\pm^{2n+1}) & =&
 \sum_{k=0}^{2n+1} C_{2 n+1}^k u^k Tr ( (P^\pm)^k (Q^\ast)^{2n+1-k}
 ) \nonumber\\
 & &= 2 \sum_{k=0}^{2n+1} C_{2n+1}^{k} u^{k}=2 (1+u)^{2n+1}
\ena where the symbol $C_n^m$ denotes $n!/m!(n-m)!$. Similarly we
have \eq Tr(\check{R}_\pm^{-2n}) =\frac 2 {(1-u)^{2n}} + \frac 2
{(1+u)^{2n}}, \qquad
 Tr(\check{R}_\pm^{-2n-1}) =\frac 2 {(1+u)^{2n+1}}.
\en The Markov traces for the links $\overline{\sigma_1^{2n}}$ and
$\overline{\sigma_1^{-2n}}$  are obtained to be \eqa
 Z(\overline{\sigma_1^{2n}}) &=& (1+u)^{-2n} Tr (\check{R}_\pm^{2n})
 =2(1+ {u^\prime}^{ n}),
 \qquad  u^\prime=(\frac {1-u} {1+u})^2, \nonumber\\
 Z(\overline{\sigma_1^{-2n}}) &=& (1+u)^{2n} Tr (\check{R}_\pm^{-2n})
 =2(1+{ u^\prime}^{- n}),
\ena  and  the Markov traces for knots $\overline{\sigma_1^{2n+1}}$
and $\overline{\sigma_1^{-2n-1}}$ are the result given by
  \eq
 Z(\overline{\sigma_1^{2n+1}})=Z(\overline{\sigma_1^{-2n-1}})=2.
\en  They detect linking numbers for links of two components and
distinguish links with nonvanishing linking numbers from knots of
one component. But they can not classify knots in the examples we
are concerned about.  In addition, we compute the Markov traces for
the Figure Eight knot
 $\sigma_1\sigma_2^{-1} \sigma_1 \sigma_2^{-1}$, the Borromean rings
 $\sigma_2 \sigma_1^{-1} \sigma_2 \sigma_1^{-1} \sigma_2 \sigma_1^{-1}$
 and the Whitehead link $\sigma_1^2 \sigma_2^{-1} \sigma_1
 \sigma_2^{-1}$, which are given by respectively
 \eq
 Z(Figure\,\, Eight)=2, \qquad Z(Borromean) =8, \qquad Z(Whitehead)=4.
 \en

Note that for simplicity we compute  the Markov trace in terms of
the $\check{R}_\pm(u)$-matrix (\ref{gate}) instead of its normalized
unitary form. To conclude this subsection, we remark that when $u=0$
the $\check{R}_\pm(u)$-matrix (\ref{gate}) is neither a universal
quantum gate nor detects the linking number, as supports the
identification of a nontrivial unitary braid representation with a
universal quantum gate.

\section{Concluding remarks and outlooks}

As a concluding remark, Figure 22, a fish diagram represents what we
have done in the whole paper under the spell of permutation and its
partial transpose. This fish sees a long history, relating the
Brauer algebra to the virtual Temperley--Lieb algebra and further to
the virtual braid group and finally to YBEs, and relating the
commutant of the orthogonal group to the virtual knot theory. She
unifies our proposal of the virtual Temperley--Lieb algebra from the
viewpoint of the virtual knot theory \cite{kauffman10, kauffman11,
kauffman12} with the Hecke algebra representation of braid groups
and link polynomials \cite{jones1, jones2, jones3} into a complete
picture.
 \begin{figure}
\begin{center}
\epsfxsize=10.5cm \epsffile{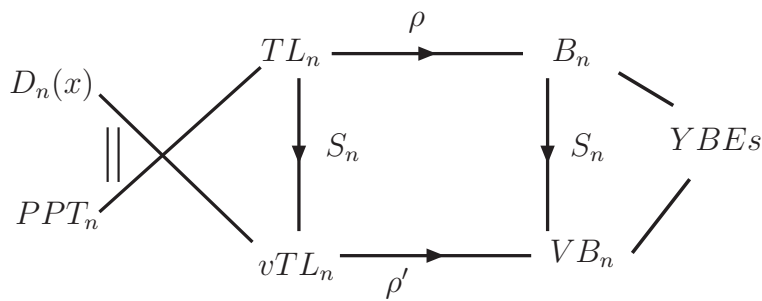}
 \caption{Fish diagram for the Brauer algebra, $vTL_n$ algebra, $VB_n$ and $YBEs$.}
 \label{fig22}
\end{center}
\end{figure}

The permutation $P$ and its partial transpose $P_\ast$ appear
similar but behave differently. The $P$ and $1\!\! 1 + u P$ are the
simplest examples for YBE solutions but $1\!\! 1 + u P$ does not
form a braid representation. The permutation's partial transpose
$P_\ast$ is an idempotent of the $TL$ algebra and the $1\!\! 1+ v
P_\ast$ can form a braid representation. It is worthwhile
emphasizing that the Werner state has the form of $1\!\! 1 + u P$
and the isotropic state has $1\!\! 1 + v P_\ast$. The YBE in terms
of matrix entries is a set of highly non-linear equations. Its
solutions are difficult to obtain unless enough constraints are
imposed. Common solutions of the braid relation (\ref{bgr}), the
multiplicative YBE (\ref{qybepl}) and additive YBE (\ref{qybead})
are found by exploring the linear combinations of $Id$, $P$ and
$P_\ast$. It is surprising because it satisfies three quite
different highly non-linear equations and roots in the existence of
flat unrestricted braid representations $FU_2$ of $P$ and $P^\ast$.

The partial transpose \cite{peres, horodecki} plays important roles
in quantum information theory. Our research is expected to be
helpful in topics such as Bell inequalities \cite{werner1}, quantum
entanglement measures \cite{werner2} and quantum data hiding
\cite{werner3}. We will apply topological contents of a family of
unitary braid representations to universal quantum gates and quantum
entanglement measures. Besides that, we set up new quantum algebras
\cite{yong1} from eight-vertex models \cite{molin1, molin2} with the
help of projectors of $P$ and $P_\ast$. In the paper, we articulate
the concepts of the $PPT_n$ algebra and virtual Temperley--Lieb
algebras. The $PPT$ algebra underlies the construction of
multipartite symmetric states \cite{eggeling} and plays crucial
roles in detecting separable quantum states \cite{werner2} and
making quantum data hiding \cite{werner3}.

The family of virtual braid representations set up representations
for the family of the virtual knot theory. The point about the
virtual knot theory is that by adding a permutation to the braiding
theory we actually bring the structure closer to quantum information
theory where the permutation (swap gate) is very important. Once one
has the swap gate one only needs to add a simple phase gate like
$Diag(1,1,1,-1)$ to obtain universality (in the presence of $U(2)$).
So it is certainly interesting to have these solutions. The
$\check{R}_\pm$-matrices (\ref{rpm}) or (\ref{baxter}) are universal
quantum gates, see \cite{molin1, molin2} for details.

Note that in \cite{yong2} we study the applications of the $TL$
algebra, Brauer algebra or virtual $TL$ algebra to quantum
teleportation phenomena. We find that the $TL$ algebra under local
unitary transformations underlies quantum information protocols
involving maximally entangled states, projective measurements and
local unitary transformations. We propose that the virtual braid
group is a natural language for the quantum teleportation.
Especially, we realize the teleportation configuration to be a basic
element of the Brauer algebra or virtual $TL$ algebra.

\section*{Acknowledgements}

 Y. Zhang thanks X.Y. Li and X.Q. Li-Jost for encouragements and supports,
 thanks M.L. Ge for fruitful collaborations and stimulating
 discussions and thanks the Mathematisches Forschungsinstitut
 Oberwolfach for the hospitality during the stay. This work is
 in part supported by NSFC--10447134  and SRF for ROCS, SEM.

For L.H. Kauffman, most of this effort was sponsored by the Defense
Advanced Research Projects Agency (DARPA) and Air Force Research
Laboratory, Air Force Materiel Command, USAF, under agreement
F30602-01-2-05022. The U.S. Government is authorized to reproduce
and distribute reprints for Government purposes notwithstanding any
copyright annotations thereon. The views and conclusions contained
herein are those of the authors and should not be interpreted as
necessarily representing the official policies or endorsements,
either expressed or implied, of the Defense Advanced Research
Projects Agency, the Air Force Research Laboratory, or the U.S.
Government. (Copyright 2006.) It gives L.H. Kauffman great pleasure
to acknowledge support from NSF Grant DMS-0245588.

\appendix

 \section{The Hecke algebra representation of the braid group}

 The Hecke algebra $H_n$ of Type $A$ is generated by $1\!\! 1$ and
 $n-1$ hermitian projections $e_i$ satisfying
 \eqa
 e_i^2 &=& e_i, \qquad (e_i)^\dag=e_i,\,\,\, i=1,\ldots,n-1, \nonumber\\
 e_i e_{i+1} e_i- \lambda e_i &=& e_{i+1} e_i e_{i+1}- \lambda e_{i+1},
 \qquad e_i e_j=e_j e_i, \,\,\, |i-j|>1.
  \ena
The parameter $\lambda$, which clearly must be in the interval
$[0,1]$ if we want a $\ast$-representation on a Hilbert space, is
fixed. Just to get three formulas used in the following, observe
 that for a pair of projections $p, q$, and a real number
 $0<\lambda<1$ the following are equivalent:
  \eq
 (1)\,\, pqp=\lambda p,\qquad
 (2)\,\, [q,pqp-\lambda p]=0, \qquad
 (3)\,\, pqp-\lambda p=qpq-\lambda q.\en

 {\bf Proof:} The turbo version is to appeal to the
universal C*-algebra generated by two projections.
(3)$\Rightarrow$(2), because the RHS commutes with $q$. Now (2)
implies that $x=pqp-\lambda p$ is in the center of the C*-algebra
generated by $p, q$. Consider any irreducible representation, in
which $x$ is then a scalar. But $xp=x$, and hence $x\neq0$ implies
$p=1$, which implies $q=\lambda $, which is not a projection. It
follows that $x=0$ in every irreducible representation, which is
(1). Finally, assuming (1), we get (2) with the roles of $p, q$
interchanged $(qpq-\lambda q)p=q(pqp-\lambda p)$. Hence by the
argument for (2)$\Rightarrow$(1) just given, both vanish.

$\hfill \Box $

Now fix $\alpha\neq\beta\in{\bf C}$, set
$$\sigma_j=\alpha e_j+\beta (1-e_j), $$
and ask, when these operators satisfy braid relations. Then, only
using $e_i^2=e_i$, but not the $TL$ relation, we get
\begin{eqnarray}
 \sigma_j\sigma_{j+1}\sigma_j-\sigma_{j+1}\sigma_j\sigma_{j+1}&=&
     \alpha\beta(\alpha-\beta)(e_j-e_{j+1})
     \nonumber\\&&\strut\quad
     +(\alpha-\beta)^3(e_je_{j+1}e_j-e_{j+1}e_je_{j+1})
\end{eqnarray}
where the terms $e_je_{j+1}$ cancel. Hence by the equivalence
(3)$\Leftrightarrow$(1), the braid relations are satisfied for the
$\sigma_i$, iff the $e_i$ satisfy the Hecke algebra of type $A$ with
\begin{equation}\label{laab}
    \lambda=\frac{-\alpha\beta}{(\alpha-\beta)^2}.
\end{equation}
Note that since the braid relation is homogenous, the expression for
$\lambda$ does not depend on a common factor of $\alpha$ and
$\beta$. Exchanging the two eigenvalues of $\sigma_i$ gives the
inverse (up to a factor), which again satisfies braid relations.
Hence the expression for $\lambda$ is symmetric in $\alpha$ and
$\beta$. Moreover, given $\lambda$, we can solve a quadratic
equation for $(\alpha/\beta)$.

Suppose $\alpha,\beta$ have the same modulus, which is equivalent to
saying that $\sigma_i$ is unitary up to a factor. Then by choosing
the factor we may set $\alpha=\exp(it)$, $\beta=\exp(-it)$, with
$0\leq t\leq\pi/2$, which produces $\lambda=1/(2\sin t)^2$. Note
that we cannot choose $t\approx0$, or, more precisely, we need $\sin
t
>1/2$, so that the eigenvalues of a unitary braid group generator
must be at least $\pi/3$ apart. At the extreme other end
($t=\pi/2\Leftrightarrow \lambda=1/4$), we can write the eigenvalues
as $\pm1$, so that we effectively have a representation of the
permutation group, rather than the braid group. In the regime
$\lambda<1/4$ we can choose both eigenvalues $\alpha,\beta$ real,
hence $\sigma_i^*=\sigma_i$. Clearly this is what happens in the
paper. Note that $e_i=\frac 1 d P_*$ satisfies the $TL_2$ algebra
with $\lambda=d^{-2}$, namely $d^3(e_je_{j+1}e_j)=de_j$ and
$d^2e_j^2=d\cdot de_j$. Then the parameters $v_\pm$ (\ref{vb}) for
fixing the braid generator correspond to the $\alpha,\beta$
satisfying (\ref{laab}).

Note on the Jones--Wenzl representation \cite{jones1, jones2,
jones3, wenzl}: Its parameter $\lambda$ denotes the quantum
factorial $[2]$ given by \eq \lambda=[2]^{-2}, \qquad
[2]=-q^{2}-q^{-2}=-2\cos\frac\pi r, \,\, q=\exp(\pi i/{2 r}), \,\,
r\ge 3. \en The number ``2" in brackets refers to $SU(2)$, while the
general $SU(k)$ theory with $r\geq k+1$ gives rise to the HOMFLY
polynomial \cite{homfly, freedman, reznikoff}. Here we have \eq
d=-q^{2}-q^{-2}, \qquad q=\pm \frac 1 2 i \sqrt{d\mp \sqrt {d^2-4}}.
\en As $r \to \infty$ and $q \to 1$, we obtain $d=2$ so that our
projector is a kind of limit of the Jones--Wenzl projector.

 \section{The proof at $d>2$ for Theorem 1 }

At $d>2$ and $a\neq 0$, Theorem 1 remarks that the $\check{R}(u)$
operator (\ref{lpa}) does not satisfy the braid relation (\ref{bgr})
and is not a solution of the YBE (\ref{qybepl}) or (\ref{qybead})
with the coefficient of $P$ as the spectral parameter. As an
example, we prove  that the $\check{R}(u)$ operator (\ref{lpa}) does
not satisfy the YBE (\ref{qybead}) for $d>2, a\neq 0$. The remaining
two statements are verified in a similar way. The left handside of
the YBE (\ref{qybead}) acts on the basis $|ijk\rangle$ in the way
 \eqa
  & & (\check{R}(u)\otimes Id)\,(Id\otimes\check{R}(u+v))\,
  (\check{R}(v)\otimes Id) |ijk\rangle  \nonumber\\
 &=& (a^3+u a v) |ijk\rangle +(a^2 v+u  a^2) |j i k \rangle
  \nonumber\\
  & & + (u a b+ 2 b a^2  +b a v+ b^2 a d+b^2 (u+v)+b^3 ) \sum_{l=1}^d
 |llk\rangle \delta_{ij} \nonumber\\
& &  +  a^2 (u+v) |ikj\rangle +a v(u+v) |j k i \rangle  +( b (u+v)
 a+  b^2 u )
 \sum_{l=1}^d |lkl\rangle \delta_{ij}  \nonumber\\
 & & + b a^2 \sum_{l=1}^d |ill\rangle\delta_{jk}+
  a v b \sum_{l=1}^d |j l l \rangle \delta_{ik}+
   (b^2 a+ b u(u+v) ) \sum^d_{l,m=1}|k l l\rangle  \delta_{ij} \nonumber\\
& &  +  u a (u+v) |kij\rangle +u v(u+v) |k j i \rangle
 + u b a \sum_{l=1}^d |l i l\rangle\delta_{jk}+
  u v b \sum_{l=1}^d |l j l \rangle \delta_{ik}
 \nonumber\\
& & + ( b a (u+v)+ b^2 v)\sum_{l=1}^d |llj\rangle \delta_{i k} + (b
v(u+v)+ a b^2   ) \sum_{l=1}^d |l l i \rangle \delta_{j k},
 \ena
while the action of the right handside of the YBE (\ref{qybead}) on
$|ijk\rangle$ has the form \eqa
 & & (Id\otimes\check{R}(v))\,(\check{R}(u+v)\otimes
  Id)\,(Id\otimes\check{R}(u))|i j k\rangle  \nonumber\\
 &=&  (a^3+ a u v ) |i j k\rangle+  a^2 (u+v) |j i k\rangle
 + u v (u+v) | k j i \rangle+ a u (u+v)| k i j \rangle
 \nonumber\\
 & & +
  (a b (u+v)+b^2  v) \sum_{l=1}^d |l i l\rangle \delta_{jk}
+ a^2 b \sum_{l}^d|l l k\rangle \delta_{ij}+
 ( a b^2+ b (u+v) v   ) \sum_{l=1}^d |l l i\rangle \delta_{jk}  \nonumber\\
  & & + a (u+v) v  |j k i\rangle +
 (a b v+a b u+ b^2 (u+v)+b^3+ 2 a^2 b+ a b^2 d  ) \sum_{l=1}^d |ill\rangle \delta_{jk}
 \nonumber\\ & &  + a^2 (u+v)  |i k j\rangle + a b v  \sum_{l}^d|l k l\rangle \delta_{ij}
  + (a b (u+v)+ u b^2 )  \sum_{l=1}^d|j l l\rangle \delta_{ki}
  \nonumber\\
 & & + (a b^2+u b (u+v)) \sum_{l=1}^d|k l l\rangle \delta_{ij}
 +u a b \sum_{l=1}^d |l l j\rangle \delta_{k i}+u  v b \sum_{l=1}^d |l j l\rangle \delta_{ki}.
 \ena
On both sides, there are $15$ terms independent of each other. The
dimension of the vector space of the three-fold tensor product is
$d^3$. If $d>2$ then $d^3>15$, we are allowed to recognize every
term on both sides and obtain three equations of $a, b, d$ given by
\eq \left\{\begin{array}{cll}
  a b u & = & a b (u+v)+b^2  v,\\
a b v & = & a b (u+v)+ b^2 u,  \\
 a^2 b & = & a b (u+v) + b^2 (u+v)+b^3+ 2 a^2 b+ a b^2 d.
\end{array}\right.
\en At $a \neq 0$ and $u, v$ as constants, there is a unique
solution: $d=2, a=-b$, supporting the statement that the
$\check{R}(u)$-matrix (\ref{matrixru}) satisfies the YBE
(\ref{qybead}).


\begin{thebibliography}{99}

  \bibitem {kauffman0} L.H. Kauffman, {\it Knots and Physics}
  (World Scientific Publishers, 2002).

 \bibitem{nielsen}
  M. Nielsen and I. Chuang, {\it Quantum Computation and Quantum Information}
 (Cambridge University Press, 1999).

  \bibitem{wolf}  M.M. Wolf, {\it Partial Transposition in Quantum Information Theory},
  PhD. Thesis (TU Braunschweig, 2003).

  \bibitem{peres} A. Peres, {\it Separability Criterion for Density Matrices},
  Phys. Rev. Lett. {\bf 77} (1996) 1413-1415. Arxiv: quant-ph/9604005.

 \bibitem{horodecki} M. Horodecki, P. Horodecki and R. Horodecki,
 {\it Separability of Mixed States: Necessary and Sufficient
 Conditions}, Phys. Lett. A {\bf 223} (1996) 1.
 Arxiv: quant-ph/9605038.

 \bibitem{werner0} R.F. Werner, {\it  Quantum Information Theory--An
  Introduction to Basic Theoretical Concepts and Experiments, Chapter Quantum
  Information Theory--An Invitation} (Springer--Verlag, New York, 2001).

 \bibitem{werner1} R.F. Werner, {\it Quantum States with Einstein-Podolsky-Rosen
   Correlations Admitting a Hidden-Variable Model},
   Phys. Rev. A {\textbf 40} (1989) 4277.

 \bibitem{yang} C.N. Yang,
 {\it Some Exact Results for the Many Body Problems in One Dimension with
 Repulsive Delta Function Interaction}, Phys. Rev. Lett. {\bf 19} (1967) 1312-1314.

 \bibitem{baxter} R.J. Baxter, {\it Partition Function of the Eight-Vertex Lattice Model},
 Annals Phys.\ {\bf 70} (1972) 193-228.

 \bibitem{horodecki1} M. Horodecki and P. Horodecki, {\it Reduction Criterion of
 Separability and Limits for A Class of Distillation Protoco},
 Phys. Rev. A {\bf 59} (1999) 4206.

 \bibitem{brauer}  R. Brauer, {\it On Algebras Which Are Connected With the
 Semisimple Continuous Groups}, Ann. of Math. {\bf 38} (1937) 857-872.

  \bibitem{werner2} K.G. H. Vollbrecht and R.F. Werner, {\it
 Entanglement Measures under Symmetry}, Phys. Rev. A {\bf 64} (2001).
 Arxiv: quant-ph/0010095.

 \bibitem{eggeling} T. Eggeling, {\it On Multipartite Symmetric States in Quantum
 Information Theory}, PhD. Thesis (TU Braunschweig, 2003).

 \bibitem{werner3} T. Eggeling and R.F. Werner, {\it Hiding Classical Data in
 Multi-Partite Quantum States},  Phys. Rev. Lett. {\bf 89} (2002).
 Arxiv: quant-ph/0203004.




 \bibitem{kauffman2}
 L.H. Kauffman and S.J. Lomonaco Jr.,  {\it Braiding Operators are
 Universal Quantum Gates},  New J. Phys. {\bf 6} (2004) 134.
 Arxiv: quant-ph/0401090.

 \bibitem{molin1} Y. Zhang, L.H. Kauffman and M.L. Ge,
 {\it Universal Quantum Gate, Yang--Baxterization and Hamiltonian}.
    Int. J. Quant. Inform., Vol. 3, {\bf 4} (2005) 669-678. Arxiv: quant-ph/0412095.

 \bibitem{molin2} Y. Zhang, L.H. Kauffman and M.L. Ge,
 {\it Yang--Baxterizations, Universal Quantum Gates and
  Hamiltonians}.  Quant. Inf. Proc. {\bf 4} (2005) 159-197. Arxiv: quant-ph/0502015.

\bibitem{kauffman3}
 L.H. Kauffman, {\it Quantum Computation and the Jones Polynomial}, in
 {\em Quantum Computation and Information}, S. Lomonaco, Jr. (ed.), AMS CONM/305, 2002,
  pp. 101-137. Arxiv: math. QA/0105255.

\bibitem {kauffman5}
 L.H. Kauffman, {\it Quantum Topology and Quantum Computing}, in
{\em Quantum Computation}, S. Lomonaco (ed.), AMS PSAPM/58, 2002,
pp. 273--303.

 \bibitem{kauffman8}
L. H. Kauffman and S. J. Lomonaco Jr., {\it Quantum Knots},
 in E. Donkor, A.R. Pirich and H.E. Brandt (eds.), Quantum Information
 and Computation II, Spie Proceedings, (12 -14 April, Orlando, FL, 2004),
 Vol. 5436, pp. 268-284. Arxiv: quant-ph/0403228.

\bibitem{kauffman4}
 L.H. Kauffman and S.J. Lomonaco Jr., {\it Quantum Entanglement and
Topological Entanglement},  New J. Phys. {\bf 4} (2002) 73.1--73.18.

 \bibitem{kauffman7}
L.H. Kauffman and S.J. Lomonaco Jr.,
 {\it Entanglement Criteria--Quantum and Topological}, in E. Donkor, A.R. Pirich
 and H.E. Brandt (eds.), Quantum Information and Computation -- Spie Proceedings,
 (21-22 April, Orlando, FL, 2003), Vol. 5105, pp. 51-58. Arxiv: quan-ph/0304091.

 \bibitem{kauffman9}
 L.H. Kauffman, {\it Teleportation Topology}.
 Opt. Spectrosc. {\bf 9} (2005) 227-232. Arxiv: quan-ph/0407224.





 \bibitem{kauffman10} L.H. Kauffman,  {\it Virtual
 Knot Theory}, European J. Comb. {\bf 20} (1999) 663-690.

 \bibitem{kauffman11} L. H. Kauffman, {\it A Survey of Virtual Knot Theory},
 in ``Proceedings of Knots in Hellas 98" (World Scientific, Singpore,  2000) 143-202.

 \bibitem{kauffman12} L. H. Kauffman, {\it Detecting Virtual Knots},
{Atti. Sem. Mat. Fis.  Univ. Modena}, Supplemento al Vol. {\bf IL}
(2001) 241-282.

  \bibitem{kamada} S. Kamada, {\it Braid Representation of Virtual Knots and Welded Knots},
 Arxiv: math. GT/0008092.

 \bibitem{lieb} H.N.V. Temperley and E.H. Lieb, {\it Relations between the `Percolation' and
  `Colouring' Problem and Other Graph-Theoretical Problems Associated with Regular Planar Lattices:
  Some Exact Results for the `Percolation' Problem}, Proc. Roy. Soc. A {\bf 322} (1971) 251-280.

 \bibitem{kauffman15} L.H. Kauffman and S.L. Lin, {\it Temperley--Lieb Recoupling Theory
  and Invariants of Three-Manifold}, Ann. of Math. Stud. {\bf 114}
  (Princeton Univ. Press, 1994).

  \bibitem{jones1} V.F.R. Jones, {\it Heck Algebra Representations of Braid Groups
  and Link Polynomials},  Ann. of Math. {\bf 126} (1987) 335-388.

  \bibitem{jones2} V.F.R. Jones, {\it Braid Groups, Heck Algebra and Type II
 Factors}, Geometric Methods in Abstract Algebras, Proc. U.S.-Japan
 Symposium (Wiley, 1986) 242-273.

 \bibitem{jones3} V.F.R. Jones, {\it A Polynomial Invariant for Knots via
  Von Neuman Algebras}, Bull. Amer. Math. Soc. (N.S.) {\bf 12} (1985) 103-111.

 \bibitem{jones}  V.F.R. Jones,  {\it Baxterization},
 Int.\ J.\ Mod.\ Phys.\ A {\bf 6} (1991) 2035-2043.

  \bibitem{murakami1} J. Murakami, {\it A State Model for the Multi-Variable Alexander
  Polynomial}, Talk at Int. Workshop on Quantum Group (Euler International Mathematical Institute,
  Leninggrad, 1990).

 \bibitem{molin4}  M.L. Ge and K. Xue, {\it Trigonometric Yang--Baxterization of Coloured
 $\check{R}$-matrix}, J. Phys. {\bf A}: Math Gen. {\bf 26} (1993) 281-291.


 \bibitem{dye1}  H.A. Dye and L.H. Kauffman, {\it Virtual Knot Diagrams and the
 Witten-Reshetikhin-Turaev Invariant}, Arxiv: math. GT/0407407.

 \bibitem{yong1} Y. Zhang, N.H. Jing and M.L. Ge,
 {\it New Quantum Algebras via RRT Relation on Eight--Vertex Models}, (in
  preparation).

 \bibitem{wu1} F.Y. Wu, {\it The Potts Model}, Rev. Mod. Phys. {\bf 54} (1982) 235-68.

  \bibitem{wu2} F.Y. Wu, {\it Knot Theory and Statistical Mechanics},
  Rev. Mod. Phys. {\bf 64} (1992) 1099-1131.

  \bibitem{kauffman14} L.H. Kauffman, {\it State Model for the Jones
 Polynomial}, Topology {\bf 26} (1987) 395-407.

 \bibitem{kulish}  P.P. Kulish,  {\it On Spin Systems Related to the Temperley--Lieb Algebra},\\
   J. Phys. {\bf A}: Math. Gen. {\bf 36} (2003) L489-L493.

  \bibitem{wenzl2} J. Birman and H. Wenzl, {\it Braids, Link Polynomials and a New
 Algebra}, Trans. Amer. Math. Soc. {\bf 313} (1989) 249-273.

 \bibitem{murakami} J. Murakami, {\it The Kauffman Polynomial of Links and Representation Theory},
  Osaka J. Math. {\bf 24} (1987) 745-758.



 \bibitem{FRR} R. Fenn, R. Rimanyi, C. Rourke, {\it The Braid Permutation Group},
{Topology} {\bf 36} (1997) 123--135.

 \bibitem{KANENOBU} T. Kanenobu, {\it Forbidden Moves Unknot a Virtual Knot},
 {J. Knot Theory and Its Ramifications} {\bf 10}  (2001) 89--96.

 \bibitem{NELSON} S. Nelson, {\it Unknotting Virtual Knots With
  Gauss Diagram Forbidden moves}, { J. Knot Theory and Its Ramifications} {\bf 10}
(2001)  931--935.

 \bibitem{kauffman13} L.H. Kauffman and S. Lambropoulou,  {\it Virtual
 Braids}, Fund. Math. {\bf 184} (2004) 159-186.
ArXiv: math.GT/0407349.

  \bibitem{molin3} M.L. Ge, L. H. Gwa and H. K. Zhao,
  {\it Yang--Baxterization of the Eight-Vertex Model: the Braid
  Group Approach}, J. Phys. {A}: Math. Gen. {\bf 23} (1990) L 795-L 798.


  \bibitem{wenzl3} H. Wenzl, {\it Representations of Heck Algebra and
  Subfactors}, PhD. Thesis (University of Pennsylvania, 1985).

  \bibitem{wenzl4} H. Wenzl, {\it  Heck Algebras of Type A and
  Subfactors}, Invent. Math. {\bf 92} (1988) 173-193.

  \bibitem{dye}
 H.A. Dye, {\it Unitary Solutions to the Yang--Baxter Equation in Dimension Four},
 Quant. Inf. Proc. {\bf 2} (2003) 117-150. Arxiv:  quant-ph/0211050.

 \bibitem{BB}
 J.L. Brylinski and R. Brylinski, {\it Universal Quantum Gates}, in {\em
Mathematics of Quantum Computation}, Chapman \& Hall/CRC Press, Boca
Raton, Florida, 2002 (edited by R. Brylinski and G. Chen).

 \bibitem{kauffman16} L.H. Kauffman, {\it Knot Diagrammatics},
 in W. Menasco and M. Thistlethwaite (eds.), {\em Handbook of Knot Theory},
 (Elsevier, 2005), pp. 233--318. Arxiv: math. GN/0410329.

  \bibitem{homfly} P. Freyd, D. Yetter, J. Hoste, W.B.R. Lickorish,
  K. Miller, and A. Ocneanu, {\it A New Polynomial Invariant of Knots and Links},
  Bull. Amer. Math. Soc. (N.S.) {\bf 12} (1985) 239-246.

 \bibitem{kauffman17} L.H. Kauffman, {\it An Invariant of Regular
  Isotopy}, Trans. Amer. Math. Monthly {\bf 95} (1988) 195-242.

  \bibitem{dye2}
   H.A. Dye and L.H. Kauffman, {\it Minimal Surface Representations of
   Virtual Knots and Links}, Arxiv: math. GT/0401035.

  \bibitem{manturov} V.O. Manture, {\it Kauffman-Like Polynomial and Curve in
  2-Surfaces}, J. Knot Theory and Its Ramifications {\bf
  12} (2003) 1145-1153.

  \bibitem{wenzl1} H. Wenzl, {\it  On the Structure of Brauer's Centralized
 Algebras}, Ann. of Math. {\bf 128} (1988) 179-193.


 \bibitem{wenzl} H. Wenzl, {\it On Sequences of Projections}, C.
  R. Math. Acad. Sci. Soc. R. Can. {\bf 9} (1987) 5-9.

   \bibitem{freedman}  M.H. Freedman, M.J. Larsen and Z. Wang,
  {\it The Two-Eigenvalue Problem and Density of Jones Representation of Braid
  Groups},  Comm. Math. Phys. {\bf 228} (2002) 177-199.

   \bibitem{reznikoff} S. Reznikoff, {\it Representations of the Temperley--Lieb Planar
  Algebra}, PhD. Thesis (University of California, Berkeley, 2002).

 \bibitem{yong2} Y. Zhang, {\it Teleportation, Braid Group and Temperleyt--Lieb
algebra}, Arxiv: quant-ph/0601050.




 \end{thebibliography}
 \end{document}